\documentclass[aps,floats]{revtex4}
\usepackage{amsmath,amssymb}
\usepackage{graphicx,epsfig}
\usepackage[greek,english]{babel}
\usepackage{bbold}

\begin{document}
\bibliographystyle {plain}

\pdfoutput=1
\def\oppropto{\mathop{\propto}} 
\def\opsimeq{\mathop{\simeq}}
\def\opoverderline{\mathop{\overline}}
\def\operarrow{\mathop{\longrightarrow}}
\def\opsim{\mathop{\sim}} 
\def\oplim{\mathop{\lim}} 

\def\opmin{\mathop{\min}} 
\def\opmax{\mathop{\max}} 

\def\fig#1#2{\includegraphics[height=#1]{#2}}
\def\figx#1#2{\includegraphics[width=#1]{#2}}


\title{ Markov trajectories : Microcanonical Ensembles based on empirical observables \\
as compared to Canonical Ensembles based on Markov generators } 


\author{ C\'ecile Monthus }
 \affiliation{Universit\'e Paris Saclay, CNRS, CEA,
 Institut de Physique Th\'eorique, 
91191 Gif-sur-Yvette, France}

\begin{abstract}
The Ensemble of trajectories $x(0 \leq t \leq T)$ produced by the Markov generator $M$ in a discrete configuration space can be considered as 'Canonical' for the following reasons : (C1) the probability of the trajectory $x(0 \leq t \leq T)$ can be rewritten as the exponential of a linear combination of its relevant empirical time-averaged observables $E_n$, where the coefficients involving the Markov generator are their fixed conjugate parameters; (C2) the large deviations properties of these empirical observables $E_n$ for large $T$ are governed by the explicit rate function $I^{[2.5]}_M  (E_.) $ at Level 2.5, while in the thermodynamic limit $T=+\infty$, they concentrate on their typical values $E_n^{typ[M]}$ determined by the Markov generator $M$. This concentration property in the thermodynamic limit $T=+\infty$ suggests to introduce the notion of the 'Microcanonical Ensemble' at Level 2.5 for stochastic trajectories $x(0 \leq t \leq T)$, where all the relevant empirical variables $E_n$ are fixed to some values $E^*_n$ and cannot fluctuate anymore for finite $T$. The goal of the present paper is to discuss its main properties : (MC1) when the long trajectory $x(0 \leq t \leq T) $ belongs the Microcanonical Ensemble with the fixed empirical observables $E_n^*$, the statistics of its subtrajectory $x(0 \leq t \leq \tau)  $ for $1 \ll \tau \ll T $ is governed by the Canonical Ensemble associated to the Markov generator $M^*$ that would make the empirical observables $E_n^*$ typical ; (MC2) in the Microcanonical Ensemble, the central role is played by the number $\Omega^{[2.5]}_T(E^*_.) $ of stochastic trajectories of duration $T$ with the given empirical observables $E^*_n$, and by the corresponding explicit Boltzmann entropy $S^{[2.5]}( E^*_. )  = [\ln \Omega^{[2.5]}_T(E^*_.)]/T  $. This general framework is applied to continuous-time Markov Jump processes and to discrete-time Markov chains with illustrative examples.

\end{abstract}

\maketitle

\section{ Introduction }

  \subsection{ Reminder on the Microcanonical Ensemble and the Canonical Ensemble for Equilibrium }
  
  \label{subsec_introEq}

The statistical physics of Equilibrium is based 
on the ergodic principle :
 the 'time average' of any observable $O\left(  {\cal C}  \right) $ of the configuration ${\cal C} $
 over the dynamical trajectory ${\cal C}(0 \leq t \leq T)  $ 
should become equivalent for large time $T \to +\infty$ to an 'Ensemble average' computed with the 
appropriate Ensemble probability $P^{Ensemble}_{eq}({\cal C})$ of the configuration ${\cal C} $
\begin{eqnarray}
\frac{1}{T} \int_0^{T} dt 
O \left(  {\cal C}(t)  \right)  
\opsimeq_{T \to \infty}
 \sum_{ {\cal C}} O \left(  {\cal C}  \right) 
P^{Ensemble}_{eq}\left({\cal C}   \right)
\label{ergodic}
\end{eqnarray} 
The various Ensembles probabilities $P^{Ensemble}_{eq}\left({\cal C}   \right) $ are adapted to the physical conditions 
that determine what observables are fixed and what observables can fluctuate. 
Let us recall two important Equilibrium Ensembles.

[MC] In the Microcanonical Ensemble for an isolated system, where the energy $E$ is fixed and cannot fluctuate,
the configurations having a different energy $E({\cal C} )\ne E$ are not possible,
while the configurations having the energy $E({\cal C} )= E$ are equiprobable
(we assume here that the configurations ${\cal C} $ and the energy $E$ are discrete, as in the Ising model for $N$ classical spins $S_i=\pm 1$ for instance)
\begin{eqnarray}
P^{Microcanonical[E]}_{eq}\left({\cal C}   \right)  
\equiv \frac{\delta_{E,E({\cal C})}}{\sum_{\cal C'} \delta_{E,E({\cal C'})}  } 
\label{peqmicro}
\end{eqnarray}

[C] In the Canonical Ensemble for a system in contact with a thermal reservoir at inverse temperature $\beta$, 
the energy $E$ can fluctuate, so that all the configurations have now a finite weight
given by  the exponential form
\begin{eqnarray}
P^{Canonical[\beta]}_{eq}({\cal C}) 
\equiv \frac{ e^{- \beta E({\cal C}) } }{ \sum_{\cal C'} e^{- \beta E({\cal C'}) } } 
\label{peqcano}
\end{eqnarray}

Despite these very important physical differences concerning the possible fluctuations of the energy $E$,
the two Ensembles become nevertheless equivalent in the thermodynamic limit of the infinite size $N \to +\infty$.
This mathematical equivalence can be understood from the physical point of view as follows :

{\it From Canonical to Microcanonical} :  If the system is in contact with a thermal reservoir and described 
by the Canonical Ensemble of Eq. \ref{peqcano}, the relative fluctuations of the energy 
with respect to the averaged energy $E^{av}_{\beta}$ in the Canonical Ensemble 
will become negligible in the thermodynamic limit $N \to +\infty$
as a consequence of the law of large numbers.
So the Canonical Ensemble at inverse temperature $\beta$
 becomes equivalent in the thermodynamic limit $N \to +\infty$
to the Microcanonical Ensemble at energy $E=E^{av}_{\beta}$.

{\it From Microcanonical to Canonical} : 
If the system of size $N$ is isolated and described by the Microcanonical Ensemble of energy $E$,
a subsystem of large size $M$ much smaller than $N$ will be described by the Canonical Ensemble
at the inverse temperature $\beta_E$ that reproduces the correct averaged intensive energy $\frac{E}{N}$.
The physical interpretation is that it is the complementary system of size $(N-M) \gg M$ that plays the role
of the thermal reservoir at the inverse temperature $\beta_E $ for the subsystem of size $M$.


  \subsection{ Analysis of non-equilibrium steady-states via time-averaging over a large-time window $T$ }

 For non-equilibrium steady states, it is natural to focus also on time-averages over long 
dynamical trajectories as in Eq. \ref{ergodic}. However, since the analog of the Equilibrium distribution 
$P^{Ensemble}_{eq}\left({\cal C}   \right) $  is not known a priori,
one introduces the time-empirical density $ \rho( {\cal C} ) $
that measures the fraction of the time spent in each configuration ${\cal C}$
\begin{eqnarray}
 \rho_T( {\cal C} ) \equiv \frac{1}{T} \int_0^T dt \ \delta_{{\cal C}(t) , {\cal C} }
\label{rhoc}
\end{eqnarray}
from which one can reconstruct the time-average of any observable $O\left(  {\cal C}  \right) $ as
\begin{eqnarray}
\frac{1}{T} \int_0^{T} dt 
O \left(  {\cal C}(t)  \right)  = \sum_{ {\cal C}} O \left(  {\cal C}  \right)  \rho_T( {\cal C} )
\label{Otimeav}
\end{eqnarray} 
The goal of the large deviation theory is then to analyze how the empirical density of Eq. \ref{rhoc}
and the time-averaged observables of Eq. \ref{Otimeav} 
fluctuate around their typical values for large time $T$ as we now recall.


  \subsection{ Large deviations for Markov trajectories over a large-time window $T$ : standard Levels 1,2,3}

 The theory of large deviations has become the unifying language for the statistical physics of equilibrium, 
 non-equilibrium and dynamical systems (see the reviews \cite{oono,ellis,review_touchette} and references therein).
It is based on the following standard classification that involves the three nested levels 
\cite{oono,ellis,review_touchette} :

(1) the Level 1 focuses on the large deviations of the time-average of Eq. \ref{Otimeav} for a given observable $O( {\cal C} )$.

(2) the Level 2 concerns the large deviations of the empirical density $\rho_T( {\cal C} ) $ of Eq. \ref{rhoc}
as a function of the configuration ${\cal C} $;
since the empirical density $\rho_T( {\cal C} ) $ allows to reconstruct the time-average of any observable $O\left(  {\cal C}  \right) $ via Eq. \ref{Otimeav}, the Level 2 can be contracted to obtain the Level 1 of any observable $O(  {\cal C}  )$.

(3) the Level 3 describes the large deviations of the whole empirical dynamical process over the time-window $T$, and can reproduce the Level 2 via contraction.

However this initial classification that was appropriate for Equilibrium 
has turned out to be inappropriate for non-equilibrium steady states for the two following reasons :

(i) in the presence of steady currents, the large deviations at Level 2 cannot be written in closed form,
while the Level 3 is actually far too general.

(ii) many interesting time-additive observables $ A$, like the currents, the activities, the entropy production, etc...
 are not of the form of the observables $O({\cal C})$ of the Level 1 
but involve also the elementary moves between two configurations, so that they cannot be reconstructed
from the empirical density $\rho_T( {\cal C} ) $ alone.

In order to overcome these difficulties, a new Level has been introduced between the Level 2 and the Level 3
and has been called 2.5 (even if it is actually much closer in spirit to the Level 2).


  \subsection{ Level 2.5 for the joint probability of the empirical density $\rho_T( {\cal C} ) $ 
  and of the empirical flows $q_T({\cal C '} , {\cal C} ) $}

 The Level 2.5 concerns
 the joint probability of the empirical density $\rho_T( {\cal C} ) $ and of the empirical flows
 $ q_T({\cal C '} , {\cal C} ) $ that measure the frequency of jumps from one configuration $ {\cal C}$ 
 to another ${\cal C '} $ configuration during the time-window $T$
 \begin{eqnarray}
 q_T({\cal C '} , {\cal C} ) \equiv \frac{1}{T}  
\sum_{ t \in [0,T] : {\cal C}(t^+) \ne {\cal C}(t^-)  } 
\delta_{{\cal C}(t^+) , {\cal C'} } \ \delta_{{\cal C}(t^-) , {\cal C} }
\label{kcc}
\end{eqnarray}

The introduction of the Level 2.5 has solved the two issues mentioned in the previous subsection as follows.

The issue (i) is now solved because
the large deviations at Level 2.5 are closed and explicit for general Markov processes,
including discrete-time Markov chains
 \cite{fortelle_thesis,fortelle_chain,review_touchette,c_largedevdisorder,c_reset,c_inference},
continuous-time Markov jump processes
\cite{fortelle_thesis,fortelle_jump,maes_canonical,maes_onandbeyond,wynants_thesis,chetrite_formal,BFG1,BFG2,verley2016,chetrite_HDR,c_ring,c_interactions,chemical,c_open,c_detailed,barato_periodic,chetrite_periodic,c_reset,c_inference,c_runandtumble,c_jumpdiff,c_skew,c_metastable,c_east,c_exclusion}
and Diffusion processes 
\cite{wynants_thesis,maes_diffusion,chetrite_formal,engel,chetrite_HDR,c_reset,c_lyapunov,c_inference,c_skew,c_metastable,coghi}. 
The Level 2 concerning the empirical density $ \rho_T( {\cal C} ) $ alone 
should be now obtained via the optimization of the Level 2.5 over the empirical flows, 
but this contraction is not always explicit.

The issue (ii) mentioned in the previous subsection is also solved because any time-additive observable 
${\cal A}[{\cal C} (0 \leq t \leq T) ]$
of the Markov trajectory ${\cal C}(0 \leq t \leq T)  $  can be rewritten 
as a linear combination of the empirical density $\rho_T( {\cal C} ) $ 
and of the empirical flows $ q_T({\cal C '} , {\cal C} ) $ with appropriate coefficients
$ f (  {\cal C}  ) $ and $g ({\cal C '} , {\cal C} ) $
 \begin{eqnarray}
{\cal A}[{\cal C} (0 \leq t \leq T) ] && =  \sum_{ {\cal C}} f (  {\cal C}  )  \rho_T( {\cal C} ) 
+ \sum_{{\cal C}} \sum_{{\cal C'} \ne {\cal C}} g ({\cal C '} , {\cal C} )q_T({\cal C '} , {\cal C} ) 
\nonumber \\
&& = \frac{1}{T} \int_0^{T} dt f \left(  {\cal C}(t)  \right)
+ \frac{1}{T}  
\sum_{ t \in [0,T] : {\cal C}(t^+) \ne {\cal C}(t^-)  } 
g\left( {\cal C}(t^+) , {\cal C}(t^-) \right)
\label{additiveconfig}
\end{eqnarray}
As a consequence, the large deviations properties of any time-additive observable ${\cal A}[{\cal C} (0 \leq t \leq T) ]$
can be derived from the Level 2.5 via the appropriate contraction.


  \subsection{ Microcanonical and Canonical Ensembles 
  associated to a given time-additive observable ${\cal A}[{\cal C} (0 \leq t \leq T) ]$  }

Since the famous Feymann-Kac formula, the standard approach to analyze the statistics of a time-additive observable ${\cal A}[{\cal C} (0 \leq t \leq T) ]$
of the Markov trajectory ${\cal C}(0 \leq t \leq T)  $ 
is based on the evaluation of the generating function of ${\cal A}[{\cal C} (0 \leq t \leq T) ]$ via 
the introduction of the appropriate deformed Markov generator.
This approach is very powerful and has been used extensively in the field of non-equilibrium
 \cite{derrida-lecture,sollich_review,lazarescu_companion,lazarescu_generic,jack_review,vivien_thesis,lecomte_chaotic,lecomte_thermo,lecomte_formalism,lecomte_glass,kristina1,kristina2,jack_ensemble,simon1,simon2,simon3,Gunter1,Gunter2,Gunter3,Gunter4,chetrite_canonical,chetrite_conditioned,chetrite_optimal,chetrite_HDR,touchette_circle,touchette_langevin,touchette_occ,touchette_occupation,derrida-conditioned,derrida-ring,bertin-conditioned,touchette-reflected,touchette-reflectedbis,c_lyapunov,previousquantum2.5doob,quantum2.5doob,quantum2.5dooblong,c_ruelle,lapolla,chabane,chabane_thesis}.
The link with the large deviations at Level 2.5 described in the previous subsection can be understood via the corresponding conditioned process obtained from the generalization of Doob's h-transform,
as explained in particular in the very detailed complementary papers \cite{chetrite_conditioned,chetrite_optimal} 
and in the Habilitation Thesis \cite{chetrite_HDR}.
Within this framework, once the trajectories probabilities ${\cal P}[{\cal C}(0 \leq t \leq T)] $
are given by the Markov model one is interested in,
one introduces for each time-additive observable ${\cal A}[{\cal C} (0 \leq t \leq T) ]$
the following Microcanonical and Canonical Ensembles
(see \cite{chetrite_canonical,chetrite_conditioned,chetrite_optimal} and references therein):
 
 [MC] the Microcanonical Ensemble associated to the fixed value $A$ of the time-additive observable ${\cal A}[{\cal C} (0 \leq t \leq T) ]$
\begin{eqnarray}
P^{Microcanonical[A]}_{[\cal A]}\left({\cal C} (0 \leq t \leq T)  \right)  
\equiv \frac{ {\cal P}[{\cal C}(0 \leq t \leq T)] \delta_{A,{\cal A} [{\cal C} (0 \leq t \leq T)]}}
{ \displaystyle \sum_{{\cal C'} (0 \leq t \leq T)} {\cal P}[{\cal C}(0 \leq t \leq T)]
\delta_{A,{\cal A} [{\cal C'} (0 \leq t \leq T)]}  } 
\label{pAmicro}
\end{eqnarray}

[C] the Canonical Ensemble associated to the parameter $k$ conjugated to the additive observable 
${\cal A}[{\cal C} (0 \leq t \leq T) ]$
\begin{eqnarray}
P^{Canonical[k]}_{[\cal A]}\left({\cal C} (0 \leq t \leq T)  \right) 
\equiv \frac{ {\cal P}[{\cal C}(0 \leq t \leq T)]  e^{ k T {\cal A}[{\cal C} (0 \leq t \leq T) ] }}
{ \displaystyle \sum_{{\cal C'} (0 \leq t \leq T)} {\cal P}[{\cal C}(0 \leq t \leq T)]
e^{k T {\cal A} {\cal C'} (0 \leq t \leq T) }} 
\label{pAcano}
\end{eqnarray}

Although the mathematical analogy with the Equilibrium Ensembles of Eqs \ref{peqmicro} and \ref{peqcano} is obvious,
there are however very important differences :

(a) Here one introduces a new Microcanonical Ensemble and a new Canonical Ensemble 
for each new time-additive observable ${\cal A}[{\cal C} (0 \leq t \leq T) ] $ one is interested in, 
while for Equilibrium, one works with the single Microcanonical Ensemble of Eq. \ref{peqmicro}
and with the single Canonical Ensemble of  Eq. \ref{peqcano} without introducing new Ensembles for each observable.

(b) For most additive observable ${\cal A}[{\cal C} (0 \leq t \leq T) ]$, 
the conjugated parameter $k$ that appear in the Canonical Ensemble of Eq. \ref{pAcano} 
remains a formal parameter 
that has no direct physical meaning and that cannot be controlled experimentally,
in contrast to the inverse temperature $\beta$ that parametrizes the Equilibrium Canonical Ensemble of Eq. \ref{peqcano}.

(c) The Ensembles of Eq. \ref{pAmicro} and Eq. \ref{pAcano} involve the trajectories probabilities 
${\cal P}[{\cal C}(0 \leq t \leq T)] $ determined by the Markov generator, while the Equilibrium Ensembles of 
Eqs  \ref{peqmicro} and Eq. \ref{peqcano} do not contain such a priori probabilities $P(\cal C)$ on their right hand-sides.
In particular, in the Microcanonical ensemble of Eq. \ref{pAmicro}, the trajectories that have the correct value
of the additive observable ${\cal A} [{\cal C} (0 \leq t \leq T)]=A $ do not have the same probability,
in constrast to the Equilibrium Microcanonical Ensemble of Eq. \ref{peqmicro},
where all the configurations ${\cal C} $ that have the correct value of the energy $E({\cal C})=E $ are equiprobable.

 
 \subsection{ Motivation and goal of the present work }
 
 The present work has been motivated by the differences (a) (b) (c) mentioned in the previous subsection.
 We will explain how a closer analogy to the Equilibrium Ensembles of Eq.  \ref{peqmicro} and  Eq. \ref{peqcano}
 can be obtained if, instead of working at the level of a single time-additive observable ${\cal A} [{\cal C} (0 \leq t \leq T)] $, 
 one works at the Level 2.5 as follows.
 
[C2.5] The trajectories probabilities ${\cal P}[{\cal C}(0 \leq t \leq T)] $ 
  determined by Markov generator $M$ can actually be considered as a 'Canonical Ensemble at Level 2.5' :
  indeed, the probability ${\cal P}[{\cal C}(0 \leq t \leq T)] $ of the trajectory ${\cal C}(0 \leq t \leq T)$ can be rewritten 
  as the exponential of a linear combination of its relevant empirical time-averaged observables $E_n$
   (namely the empirical density and the empirical flows), 
   where the coefficients involving the Markov generator are their fixed conjugate parameters.
   
[MC2.5]   It is then natural to introduce also the corresponding 'Microcanonical Ensemble at Level 2.5',
 where all the relevant empirical variables $E_n$ (namely the empirical density and the empirical flows)
 are fixed to some values $E^*_n$ 
 and cannot fluctuate anymore for finite $T$ : 
 the trajectories ${\cal C}(0 \leq t \leq T) $ that have not the correct values $E_n^*$ have zero weight,
 while all the trajectories ${\cal C}(0 \leq t \leq T) $ that have the correct values $E_n^*$ are equiprobable.
 In this Microcanonical Ensemble at Level 2.5, the central role is thus played by the number $\Omega^{[2.5]}_T(E^*_.) $ of the stochastic trajectories of duration $T$ with the given empirical observables $E^*_n$, and by the corresponding explicit Boltzmann entropy $S^{[2.5]}( E^*_. )  = [\ln \Omega^{[2.5]}_T(E^*_.)]/T  $.  
 
 The goal of the present paper is thus to describe in detail the properties of these Canonical and Microcanonical 
 Ensembles at Level 2.5.

 \subsection{ Organization of the paper }

The paper is organized as follows.
In section \ref{sec_general}, we explain in detail the general properties of the Canonical and the Microcanonical Ensembles at Level 2.5 with their links. 
This framework is then applied to continuous-time Markov Jump processes in section \ref{sec_jump}.
In section \ref{sec_EntropyJump}, we focus on the special case of undirected Markov Jump processes (where the jumps between two configurations are either both possible or both impossible) in order to describe
how the entropy $S^{[2.5]}( E_. ) $ at Level 2.5 as a function of all the relevant empirical observables $E_n$ can be contracted to obtain the explicit entropies of many other lower levels.
Our conclusions are summarized in section \ref{sec_conclusion}.
In Appendix \ref{app_chain}, the general framework of section \ref{sec_general}
is applied to discrete-time Markov chains.


\section{ Canonical and Microcanonical Ensembles for Markov trajectories    }

\label{sec_general}

In this section, we outline the general principles before the specific applications to continuous-time Markov Jump processes in section \ref{sec_jump} and to discrete-time Markov chains in Appendix \ref{app_chain}.

\subsection{ Canonical Ensemble of trajectories $x(0 \leq t \leq T)$ associated to the Markov generator $M$   }

Let us consider a Markov process in a discrete configuration space
converging towards some normalizable steady state.
Since the Markov generator $M$ defines the dynamical rules,
its matrix elements can be considered as fixed 'intensive variables'
that will govern the statistics of the trajectories $x(0 \leq t \leq T)$
with probabilities $ {\cal P}_M[x(0 \leq t \leq T)] $ normalized to unity
\begin{eqnarray}
1= \sum_{x(0 \leq t \leq T)}  {\cal P}_M[x(0 \leq t \leq T)] \equiv  \sum_{x(.)}  {\cal P}_M[x(.)]
\label{normaptraj}
\end{eqnarray}
where the last simplified notations will be used in order to ease the read of equations.

\subsubsection{ Identification of the relevant empirical observables $E_n \left[ x(.) \right] $ that determine the trajectories probabilities ${\cal P}_M\left[ x(.) \right] $ }

The information per unit time ${\cal I}_M\left[ x(0 \leq t \leq T) \right] $ of a given trajectory $x(0 \leq t \leq T)$
of probability ${\cal P}_M\left[ x(.) \right] $
\begin{eqnarray}
{\cal I}_M\left[ x(.) \right]  \equiv - \frac{ \ln \left( {\cal P}_M\left[ x(.) \right] \right) }{T}
\label{informationdef}
\end{eqnarray}
depends on the trajectory $x(0 \leq t \leq T) $  
\begin{eqnarray}
{\cal I}_M\left[ x(.) \right]  = I_M \left(E_. \left[ x(.) \right]  \right)
\label{informationempi}
\end{eqnarray}
only via a collection 
of relevant empirical observables $E_n \left[ x(.) \right]  $ labelled by $n$ :
they correspond to the time-averaged density and to the time-averaged flows
during the time window $T$ (see Eqs \ref{rho1pj} and \ref{jumpempiricaldensity} for continuous-time jump processes
 as well as Eq. \ref{rho2pt} for discrete-time Markov chains).


\subsubsection{ The information $ I_M \left(E_.  \right) $ as a linear combination of the relevant empirical observables $E_n$}

As a consequence of the Markov property that allows to decompose the trajectory probability via the Chapman-Kolmogorov Equation,
the information $ I_M \left(E_.  \right) $ is actually 
a linear combination of these relevant empirical time-averaged observables $E_n$
\begin{eqnarray}
 I_M \left(E_.  \right) = \sum_n E_n i_n(M)
\label{informationlinearity}
\end{eqnarray}
while the coefficients
\begin{eqnarray}
 i_n(M) \equiv \frac{ \partial  I_M \left(E_.  \right) }{\partial E_n} 
\label{informationlinearityderi}
\end{eqnarray}
 involve the matrix elements of the Markov generator $M$ in a very simple way 
 (see Eqs \ref{informationempimaster} and \ref{lambdajump} for continuous-time jump processes
 as well as Eq \ref{actionchain} and \ref{lambdachain} for discrete-time Markov chains).


\subsubsection{ Trajectory probability in terms of the relevant empirical observables $ E_. \left[ x(.) \right]=E_.$}

In summary, the probability $ {\cal P}_M[x(.)]  $ of the trajectory $x(0 \leq t \leq T)$ 
can be rewritten in terms of its relevant empirical observables $ E_n \left[ x(.) \right]$ as
\begin{eqnarray}
  {\cal P}_M[x(.)] 
&&  = e^{\displaystyle  -T   {\cal I}_M \left[ x(.) \right]    }  
 = e^{\displaystyle  -T  I_M \left(E_. \left[ x(.) \right]  \right)     } 
 =    \sum_{E_.} 
  \left[ \prod_n \delta \left( E_n - E_n \left[ x(.) \right] \right) \right] 
 e^{\displaystyle  -T    I_M \left( E_. \right) }
\nonumber \\
&&  =   \sum_{E_.} 
  \left[ \prod_n \delta \left( E_n - E_n \left[ x(.) \right] \right) \right] 
 e^{\displaystyle  -T     \sum_n E_n i_n(M)
  }
 \label{ptrajinfoempi}
\end{eqnarray}
where the exponential involves a linear combination of the relevant empirical time-averaged observables $E_n$, while the coefficients $i_n(M)$ determined by the Markov generator $M$ can be considered as their fixed conjugate parameters.


\subsubsection{ Number $\Omega^{[2.5]}_T ( E_. ) $ of trajectories $x(0 \leq t \leq T)$ with the same relevant empirical observables $ E_. \left[ x(.) \right]=E_.$}

\label{subsec_omega}

An important consequence of Eq. \ref{ptrajinfoempi}
is that all the trajectories $ x(0 \leq t \leq T)  $
 that have the same relevant empirical observables $E_n=E_n [x(0 \leq t \leq T)] $
  have the same probability.
  The normalization over all possible trajectories of Eq. \ref{normaptraj}
can be thus rewritten as a sum over these relevant empirical observables $E_n$
\begin{eqnarray}
1= \sum_{x(0 \leq t \leq T)}  {\cal P}[x(0 \leq t \leq T)] 
=
 \sum_{E_.} \Omega^{[2.5]}_T(E_. ) e^{\displaystyle  -T  I_M \left( E_.  \right) }
\label{normaempi}
\end{eqnarray}
where the number $  \Omega^{[2.5]}_T(E_. )$ of trajectories of duration $T$ 
associated to given values $E_n $ of these 
empirical observables
\begin{eqnarray}
\Omega^{[2.5]}_T ( E_. ) \equiv \sum_{x(0 \leq t \leq T)}  \left[ \prod_n \delta \left( E_n - E_n \left[ x(.) \right] \right) \right] 
\label{omegaempi}
\end{eqnarray}
  will grow exponentially with respect to the time-window $T$ in the limit $T \to +\infty$ 
\begin{eqnarray}
 \Omega^{[2.5]}_T( E_.) \opsimeq_{T \to +\infty} C^{[2.5]}(E_.) \ e^{\displaystyle T S^{[2.5]}( E_. )  }
\label{omegat}
\end{eqnarray}
The prefactor $C^{[2.5]}(E_.)$ denotes the appropriate constitutive constraints for the empirical observables $E_n$
(see Eq. \ref{constraints2.5master} for continuous-time jump processes
 as well as Eq. \ref{constraints2.5chain} for discrete-time Markov chains).
The function $S^{[2.5]}( E_. )  = \frac{\ln \Omega^{[2.5]}_T(E_.) }{ T }  $ represents the 
Boltzmann intensive entropy of the set of trajectories of duration $T$ with given intensive empirical observables $E_n $.
Note that since the Boltzmann entropy is the cornerstone of the statistical physics of Equilibrium,
the definition of Boltzmann entropies for non-equilibrium systems has motivated a lot of various studies 
(see for instance \cite{Gold_Leb} and references therein).


\subsubsection{ Boltzmann intensive entropy $S^{[2.5]}( E_. ) $ as a function of the relevant empirical observables $E_n $}

The entropy $S^{[2.5]}( E_. ) $ introduced in Eq. \ref{omegat}
can be evaluated without any combinatorial computation as follows.
The normalization of Eq. \ref{normaempi} becomes for large $T$
\begin{eqnarray}
1  \opsimeq_{T \to +\infty} \sum_{E_.} C^{[2.5]}(E_.) \  e^{\displaystyle  T \left[ S^{[2.5]}( E_.  ) - I_M \left( E_. \right)   \right] }
\label{normaempit}
\end{eqnarray}
When the empirical variables $E_n $ take their typical values $E_n^{typ[M]}$ for the Markov generator $M$,
the exponential behavior in $T$ of Eq. \ref{normaempit}
should exactly vanish,
i.e. the entropy $S^{[2.5]}( E_.^{typ[M]}  ) $ should exactly compensate the information $I_M \left( E_.^{typ[M]}\right)   $ 
\begin{eqnarray}
S^{[2.5]}( E_.^{typ[M]}  )=    I_M \left( E_.^{typ[M]}\right)  = \sum_n E_n^{typ[M]} i_n(M)
\label{compensation}
\end{eqnarray}
To obtain the intensive entropy $S^{[2.5]}( E_.  ) $ for other given values $E_n  $ of the empirical observables,
one just needs to introduce the modified Markov generator $M^{E_.}$ that would make 
the empirical values $E_n $ typical for this modified Markov generator
\begin{eqnarray}
   E_n = E_n^{typ[M^{E_.}]}
\label{modeleeff}
\end{eqnarray}
and to use Eq. \ref{compensation} for this modified Markov generator $M^{E_.}$ to obtain
the entropy $S^{[2.5]}( E_.  ) $ as a function of $E_.$
\begin{eqnarray}
S^{[2.5]}( E_.) = S^{[2.5]}( E_.^{typ[M^{E_.}]}  )=    I_{M^{E_.}} \left( E^{typ[M^{E_.}]}\right)   =   I_{M^{E_.}} \left( E_.\right)  
\label{entropyempi}
\end{eqnarray}
Plugging the linear expression of Eq. \ref{informationlinearity} for the information 
into Eq. \ref{entropyempi} yields that the entropy
\begin{eqnarray}
 S^{[2.5]}( E_.)= \sum_n E_n i_n(M^{E_.})
\label{entropynonlinear}
\end{eqnarray}
is nonlinear with respect to the empirical observables $E_n$,
since the empirical observables $E_.$ appear in the modified Markov generator
 $M^{E_.} $ involved in the coefficients $ i_n(M^{E_.}) $.

 However, the linearity of the information of Eq. \ref{informationlinearity}
 can be used to write the coefficients of Eq. \ref{informationlinearityderi}
 for the special values $E_n^{typ[M]} $ of the empirical observables
 in order to use Eq. \ref{compensation}
\begin{eqnarray}
 i_n(M)  
 = \frac{ \partial  I_M \left(E_.^{typ[M]} \right) }{\partial E_n^{typ[M]}} = \frac{ \partial  S^{[2.5]}\left(E_.^{typ[M]} \right) }{\partial E_n^{typ[M]}}
\label{informationlinearityderityp}
\end{eqnarray}
that can be now rewritten for the modified Markov generator $M^{E_.}$ that 
make the empirical values $E_n = E_n^{typ[M^{E_.}]}$ typical (Eq. \ref{modeleeff})
to obtain that the derivatives of the entropy $S^{[2.5]}\left(E_. \right) $ with respect to the empirical observables $E_n$ 
\begin{eqnarray}
\frac{ \partial  S^{[2.5]} \left(E_. \right) }{\partial E_n} = i_n(M^{E_.})  
\label{derientropyempirical}
\end{eqnarray}
 are given by the coefficients $i_n(M^{E_.})  $ associated to the modified Markov generator $M^{E_.} $.
The comparison with the direct derivative of Eq. \ref{entropynonlinear}
 \begin{eqnarray}
\frac{ \partial  S^{[2.5]} \left(E_. \right) }{\partial E_n}= i_n(M^{E_.})   + \sum_m E_m \frac{ \partial  i_m(M^{E_.}) }{\partial E_n}
\label{entropynonlinearderi}
\end{eqnarray}
yields that Eq. \ref{derientropyempirical} requires the vanishing of the second contribition of Eq. \ref{entropynonlinearderi}
for any $n$
 \begin{eqnarray}
 \sum_m E_m \frac{ \partial  i_m(M^{E_.}) }{\partial E_n} =0
\label{identityzero}
\end{eqnarray}
The property of Eq. \ref{derientropyempirical} can be explicitly checked in Eq. \ref{entropyempijumpderi}
for continuous-time jump processes
 and in Eq. \ref{entropyempichainderi} for discrete-time Markov chains.


\subsubsection{ Large deviations at Level 2.5 for the relevant empirical observables $E_.$}

 Eq \ref{normaempi} can be interpreted as the normalization
 \begin{eqnarray}
 1  = \sum_{ E_. }  P^{[2.5]}_M  (E_.)
\label{normaprobaempi}
\end{eqnarray}
 for the probability $ P^{[2.5]}_M (E_.) $ to see the relevant empirical observables $E_n$ 
over the set of trajectories of duration $T$ generated with the Markov generator $M$.
For large time $T \to +\infty$,  the asymptotic behavior of $\Omega^{[2.5]}_T( E_.) $ in Eq.
\ref{omegat} can be plugged into Eq. \ref{normaempi}
to obtain the large deviation properties for all the relevant empirical observables $E_n$ 
that determine the trajectories probabilities
\begin{eqnarray}
P^{[2.5]}_M  (E_.) \opsimeq_{T \to +\infty}  C^{[2.5]}(E_.) \  e^{\displaystyle  T \left[ S^{[2.5]}(E_.) -  I_M(E_.)  \right] }   
= C^{[2.5]}(E_.) \ e^{\displaystyle  - T  I^{[2.5]}_M  (E_.)    }
\label{level2.5}
\end{eqnarray}
where the positive rate function at Level 2.5
\begin{eqnarray}
I^{[2.5]}_M  (E_.)  = I_M(E_.)  - S^{[2.5]}( E_.  )  =  I_M(E_.) -  I_{M^{E_.}} \left( E_.\right)  \geq 0
\label{rate2.5}
\end{eqnarray}
is simply given by the difference between
 the information $I_M  (E_.) $ associated to the Markov generator $M$
and the information $I_{M^{E_.}}   (E_.) $ associated to the modified Markov generator $M^{E_.}$ that would make the empirical values $E_n$ typical (see Eq. \ref{modeleeff}).

The typical values $E_n^{typ[M]} $ for the Markov generator $M$
are the only values of the empirical observables $E_n$
that satisfy the constitutive constraints $C^{[2.5]}(E_.)$
and that make the rate function of Eq. \ref{rate2.5} vanish
\begin{eqnarray}
I^{[2.5]}_M  (E_.^{typ[M]})  = 0 
\label{rate2.5vanish}
\end{eqnarray}
For large $T$, the Level 2.5 of Eq. \ref{level2.5} describes how rare it is to see 
empirical observables $E_.$ different from their typical values $E_.^{typ[M]}$,
while in the thermodynamic limit $T=+\infty$, the Level 2.5 of Eq. \ref{level2.5master} reduces 
to delta functions for all the relevant empirical observables $E_n$ that determine the trajectories probabilities
\begin{eqnarray}
P^{[2.5]}_M(E_.) \opsimeq_{T = +\infty} \prod_n \delta \left (E_n- E_n^{typ[M]} \right)
\label{level2.5infinity}
\end{eqnarray}


\subsubsection{ Contraction of the explicit Level 2.5 to obtain large deviations properties of all the lower Levels   }

\label{subsec_contractionLD}

As recalled in the Introduction, the explicit large deviations at Level 2.5 of Eq. \ref{level2.5}
concerning the joint distribution of all the relevant empirical observables $E_n$
allows to analyze via contraction all the large deviations properties of lower levels.
Let us mention some important examples :

(i) The joint distribution of any subset of the relevant empirical observables $E_n$
can be obtained via the integration of Eq. \ref{level2.5}
over all the empirical observables that one does not wish to keep.
For instance, the distribution of the empirical density $\rho_.$ alone
can be obtained via the integration of Eq. \ref{level2.5} over all the empirical flows
with their constitutive constraints
to obtain the large deviations at Level 2
\begin{eqnarray}
P^{[2]}_M  (\rho_.) \opsimeq_{T \to +\infty}   C^{[2]}(\rho_.) \ e^{\displaystyle  - T  I^{[2]}_M  (\rho_.)    }
\label{level2}
\end{eqnarray}
where the constitutive constraint $C^{[2]}(\rho_.) $ corresponds to the normalization
of the empirical density $\rho_.$.
However the rate function $ I^{[2]}_M  (\rho_.)  $ at Level 2 for the empirical density alone 
$\rho_.$ is not always explicit when the Markov generator $M$ does not correspond to some detailed-balance equilibrium dynamics.

(ii) Any intensive time-additive observable ${\cal A}_M\left[ x(.) \right]$
of the trajectory $x(0 \leq t \leq T) $ 
can be rewritten as a linear combination of the empirical observables $E_n\left[ x(.) \right]$ 
with appropriate coefficients $a_n(M)$ that may depend on the Markov generator $M$
\begin{eqnarray}
{\cal A}_M\left[ x(.) \right]  = \sum_n E_n \left[ x(.) \right]  a_n(M) \equiv A_M \left( E_.\left[ x(.) \right]\right)
\label{additive}
\end{eqnarray}
Since all the trajectories $ x(0 \leq t \leq T)  $
 that have the same empirical observables $E_n=E_n [x(0 \leq t \leq T)] $
  have the same value for the time-additive observable of Eq. \ref{additive},
 its probability distribution can be evaluated from the Level 2.5 of Eq. \ref{level2.5}
\begin{eqnarray}
P^{[Add]}_M(A)  && \equiv  \sum_{x(0 \leq t \leq T)}  {\cal P}[x(0 \leq t \leq T)] 
\delta \left(A - \sum_n E_n \left[ x(.) \right]  a_n(M) \right)
 = \sum_{ E_. }  P^{[2.5]}_M  (E_.) \delta \left(A - \sum_n E_n  a_n(M) \right)
\nonumber \\
&& \opsimeq_{T \to +\infty}  \sum_{ E_. }  C^{[2.5]}(E_.) \ e^{\displaystyle  - T  I^{[2.5]}_M  (E_.)    }
 \delta \left(A - \sum_n E_n  a_n(M) \right) 
 \opsimeq_{T \to +\infty}  e^{ -T I^{[Add]}_M(A) }
\label{additiveproba}
\end{eqnarray}
The rate function $I^{[Add]}_M(A)  $ for the time-additive observable 
corresponds to the minimization of the rate function $I^{[2.5]}_M  (E_.)  $
at Level 2.5
over the empirical observables $E_n$ satisfying the constitutive constraints $C^{[2.5]}(E_.) $
and the supplementary constraint $A = \sum_n E_n  a_n(M) $ reproducing the correct value $A $ of the additive observable
\begin{eqnarray}
I^{[Add]}_M(A)   = \opmin_{   
\substack{E_. :  C^{[2.5]}(E_.) {\rm and } \\ \sum_n E_n  a_n(M)=A}}
  \left(  I^{[2.5]}_M  (E_.)  \right)
\label{additivecontraction2.5}
\end{eqnarray}
 This contraction can be analyzed via the method of Lagrange multipliers to impose the constraints,
 but again it is not always explicit. 
 When the additive observable is the information ${\cal I}_M\left[ x(.) \right]  $ of Eqs \ref{informationempi} and \ref{informationlinearity}, its large deviations properties 
 can be analyzed via the Ruelle thermodynamic formalism recently revisited 
 in \cite{c_ruelle}.
 Besides this example of the information ${\cal I}_M\left[ x(.) \right]  $, there are 
 of course many other interesting time-additive observables depending on the 
specific models.

(iii)  The Level 1 concerns the additive observables ${\cal A}_M\left[ x(.) \right]  $
of Eq. \ref{additive} that can be reconstructed from
the empirical density $\rho_.$ alone (i.e. that do not involve the empirical flows)
and whose large deviations can be thus obtained from the contraction of the Level 2 of Eq. \ref{level2}.


\subsubsection{ Relation with the Kolmogorov-Sinai entropy $h_M^{KS}$ of Markov trajectories}

The Kolmogorov-Sinai entropy $h_M^{KS}$ is defined 
via the averaged value of the intensive information
${\cal I}_M[x(0 \leq t \leq T)]  $ of Eq. \ref{informationdef}
over the trajectories $x(0 \leq t \leq T) $ drawn with their probabilities ${\cal P }_M\left[ x(0 \leq t \leq T) \right]   $
\begin{eqnarray}
h^{KS}_M && \equiv  \oplim_{T \to +\infty} \left(  \sum_{x(.)}   {\cal P}_M[x(.)]  {\cal I}_M[x(.)]  \right)  
\nonumber \\
&&  = \oplim_{T \to +\infty} 
\left( - \frac{1}{T}  \sum_{x(0 \leq t \leq T)}   {\cal P}_M[x(0 \leq t \leq T)] 
\ln \left( {\cal P}_M[x(0 \leq t \leq T)] \right)\right)
\label{hksdef}
\end{eqnarray}
So it characterizes the linear growth in $T$ of the 
Shannon entropy $ {\cal S}^{Shannon}_M(T)$ associated to the probability distribution 
${\cal P}_M[x(0 \leq t \leq T)] $ of the trajectories
\begin{eqnarray}
{\cal S}^{Shannon}_M(T) \equiv  - \sum_{x(0 \leq t \leq T)}   {\cal P}_M[x(0 \leq t \leq T)] 
\ln \left( {\cal P}_M[x(0 \leq t \leq T)] \right) \oppropto_{T \to +\infty} T \ h^{KS}_M
\label{defdynentropy}
\end{eqnarray}

The average of ${\cal I}_M[x(0 \leq t \leq T)]  $ over trajectories of Eq. \ref{hksdef} can be rewritten as an average 
of the information $  I_M ( E_.)$ of Eq. \ref{informationempi}
over the probability $P^{[2.5]}_M (E_.) $ of Eq. \ref{level2.5}
for the empirical observables $E_n$ 
\begin{eqnarray}
h^{KS}_M =  \oplim_{T \to +\infty} \left(  \sum_{E_.} 
P^{[2.5]}_M (E_.)  I_M (E_.)
  \right)  
\label{hksempi}
\end{eqnarray}
For $T=+\infty$, only the typical values
$E_.^{typ[M]}$ of the empirical observables survive in Eq. \ref{level2.5infinity}
so that the Kolmogorov-Sinai entropy $h^{KS}_M$ 
reduces to the information $ I_M \left( E_.^{typ[M]} \right) $ associated to the typical values
$E^{typ[M]}$ of the empirical observables
\begin{eqnarray}
h^{KS}_M  =  I_M \left( E_.^{typ[M]}  \right)
\label{hksempityp}
\end{eqnarray}
Eq. \ref{compensation} yields that the Kolmogorov-Sinai entropy $h^{KS}_M$
also coincides with
the entropy $S^{[2.5]}( E_.^{typ[M]}  ) $  associated to the typical values
$E^{typ[M]}$ of the empirical observables
\begin{eqnarray}
h^{KS}_M  = S^{[2.5]}( E_.^{typ[M]}  )
\label{hksempitypSalso}
\end{eqnarray}

The physical meaning is that the average over trajectories with their probabilities ${\cal P}[x(0 \leq t \leq T)] $
 is actually dominated in the thermodynamic limit $T \to +\infty$
 by the number of Eq. \ref{omegat} 
 \begin{eqnarray}
\Omega^{[2.5]}_T(E_.^{typ[M]} ) \opsimeq_{T \to +\infty}  e^{T S^{[2.5]}( E_.^{typ[M]}  ) } = e^{T h_M^{KS} }
\label{omegaks}
\end{eqnarray}
   of trajectories corresponding to the information of Eq. \ref{hksempityp}
  \begin{eqnarray}
  I_M \left( E_.^{typ[M]}  \right) = h^{KS}_M
\label{infoks}
\end{eqnarray} 
  so that all these trajectories have the same probability given by  
  $ e^{-T h_M^{KS}}= \frac{1}{\Omega^{[2.5]}_T(E_.^{typ[M]} )}$.
As a consequence, it is interesting to introduce the corresponding notion of 
Microcanonical Ensemble described in the next subsection.


\subsection{ Microcanonical Ensemble at Level 2.5 based on fixed relevant empirical observables $E_n^*$ }

\subsubsection{ Microcanonical Ensemble at Level 2.5 where the relevant empirical observables $E_n^*$ cannot fluctuate}

The concentration property of Eq. \ref{level2.5infinity} for the all the relevant 
empirical observables in the thermodynamic limit $T=+\infty$
and the discussion around Eqs \ref{omegaks} and \ref{infoks}
suggests to introduce the notion of the 'Microcanonical Ensemble' at Level 2.5
where the all the relevant empirical variables $E_n$
 are fixed to some values $E^*_n$
  satisfying the constitutive constraints $C^{[2.5]}(E^*_.)$
 \begin{eqnarray}
P^{Micro[2.5]}_{E^*_.} (E_.)  = \prod_n \delta(E_n-E_n^*)
\label{microcanodeltaempirical}
\end{eqnarray}
and thus cannot fluctuate for finite $T$ in contrast to the Canonical fluctuations at Level 2.5 of
Eq. \ref{level2.5} associated to the Markov generator $M$.

Note that here the possible elementary moves of the trajectories correspond to 
the empirical flows that are fixed to non-vanishing values,
while the empirical flows that are fixed to vanishing values correspond to impossible elementary moves.


\subsubsection{ Probabilities of trajectories $x(0 \leq t \leq T) $ in the Microcanonical Ensemble at Level 2.5 }

In the Microcanonical Ensemble of Eq. \ref{microcanodeltaempirical}, 
only the trajectories $x(0 \leq t \leq T)  $
corresponding to the given empirical values $E^*_n$ have a non-zero weight,
and all these allowed trajectories have the same weight given by the inverse of their number $\Omega^{[2.5]}_T(E_.^*) $ of Eq. \ref{omegaempi}
\begin{eqnarray}
{\cal P}^{Micro[2.5]}_{E^*_.}[x(.)]   
 =  \frac{ 1}{\Omega^{[2.5]}_T (E_.^*) } \prod_n \delta\left(E_n-E_n^*[x(0 \leq t \leq T)  ]\right)
\label{ptrajmicro}
\end{eqnarray}
Using the asymptotic behavior for large $T$ of Eq. \ref{omegat}, 
the trajectory propability of Eq. \ref{ptrajmicro} involves the entropy $S^{[2.5]}(E_.^*)$ 
\begin{eqnarray}
{\cal P}^{Micro[2.5]}_{E_.^*}[x(.)]   
 \opsimeq_{T \to +\infty} 
 e^{\displaystyle - T  S^{[2.5]}(E_.^*)  }
\prod_n \delta \left(E_n^*- E_n[x(0 \leq t \leq T)  ] \right)
\label{ptrajmicrot}
\end{eqnarray}


\subsubsection{ Statistics of the subtrajectories on $[0,\tau]$ of the Microcanonical Ensemble trajectories on $[0,T]$ for $1 \ll \tau \ll T$ }

\label{subsec_subtraj}

In the subsection \ref{subsec_omega}, we have seen 
how the analysis of the Canonical Ensemble associated to the Markov generator $M$
actually involves the Microcanonical Ensemble via the number $\Omega^{[2.5]}_T(E_.) $ 
of trajectories with given empirical observables $E_n$.
In the present subsection, we would like to see if the Canonical Ensemble 
can emerge to describe the statistics of the long subtrajectory $x(0 \leq t \leq \tau)$
belonging to the much longer trajectory $x(0 \leq t \leq T) $ drawn with the Microcanonical distribution of Eq. \ref{ptrajmicrot}, i.e. we are interested in the regime
\begin{eqnarray}
1 \ll \tau \ll T
 \label{subtrajregime}
\end{eqnarray}
Since this property would be the direct analog of the well-know property for Equilibrium Ensembles 
(as recalled at the end of the subsection \ref{subsec_introEq} of the Introduction), 
it is interesting to try to translate step by step
 the standard derivation for Equilibrium Ensembles which is based on the Taylor expansion of the Boltzmann entropy
 (see your favorite textbook on the statistical physics of Equilibrium).

When the very long trajectory $x(0 \leq t \leq T) $ with the fixed empirical observables $E_n^*$
is decomposed into the long subtrajectory $x(0 \leq t \leq \tau)$ of empirical observables $E_n$
and its much longer complementary subtrajectory $x(\tau \leq t \leq T)$ of empirical observables $\hat E_n$,
we can use the fact that the empirical observables $E_n$ are given by time-averaged properties
to write
\begin{eqnarray}
 T E_n^* = \tau E_n + (T-\tau) \hat E_n
 \label{esub}
\end{eqnarray}
The number $ \Omega^{[2.5]}_T(E_.^*) $ of total trajectories
can be thus computed from the numbers $ \Omega^{[2.5]}_{\tau} (E) $ and $\Omega^{[2.5]}_{T-\tau} (\hat E ) $ of the two subtrajectories via the multidimensional convolution
\begin{eqnarray}
 \Omega^{[2.5]}_T(E_.^*) && = \sum_{E_.} \sum_{\hat E_. } 
 \Omega^{[2.5]}_{\tau} (E_.) \Omega^{[2.5]}_{T-\tau} (\hat E_. ) 
 \prod_n \delta \bigg( E_n^* - \frac{\tau}{T} E_n - \left( 1- \frac{\tau}{T}\right) \hat E_n  \bigg)
  \nonumber \\
&& =  \sum_{E_.} \sum_{\hat E_. } 
 \Omega^{[2.5]}_{\tau} (E_.) \Omega^{[2.5]}_{T-\tau} (\hat E_. ) 
 \prod_n \frac{ \delta \bigg(  \hat E_n - \frac{E_n^* - \frac{\tau}{T} E_n }{ 1- \frac{\tau}{T}}  \bigg) }{ 1- \frac{\tau}{T}} 
 \nonumber \\
&& \opsimeq_{\frac{\tau}{T} \ll 1} \sum_{E_.} 
 \Omega^{[2.5]}_{\tau} (E_.) \Omega^{[2.5]}_{T-\tau} 
  \left( \frac{E_.^* - \frac{\tau}{T} E_. }{ 1- \frac{\tau}{T}}  \right) 
\label{omegaconvol}
\end{eqnarray}
As a consequence, the probability $p_{\tau}(E_.)$ to see the empirical observables $E_n$
during the long subtrajectory $x(0 \leq t \leq \tau)$ of duration $\tau$ reads
\begin{eqnarray}
p_{\tau}(E_.)  \simeq \Omega^{[2.5]}_{\tau} (E_.) \ 
\frac{ \Omega^{[2.5]}_{T-\tau}   \left( \frac{E_.^* - \frac{\tau}{T} E_. }{ 1- \frac{\tau}{T}}  \right) }
 { \Omega^{[2.5]}_T(E_.^*) } 
\label{ptauE}
\end{eqnarray}
Since the constraints $C^{[2.5]}(E_.^*)$ are satisfied from the definition of the Microcanonical Ensemble,
let us assume that the constraints $C^{[2.5]}( \hat E_.)$ are satisfied whenever the constraints $C^{[2.5]}(E_.)$
are satisfied, as can be checked on explicit examples.
Using the asymptotic form of Eq. \ref{omegat} for large $\tau$, for large $(T-\tau)$ and for large $T$,
Eq. \ref{ptauE}
becomes
\begin{eqnarray}
p_{\tau}(E_.)  \simeq C^{[2.5]}(E_.) 
 e^{ \tau S^{[2.5]}(E_.) - T S^{[2.5]}(E_.^*) + (T-\tau) S^{[2.5]}  \left( \frac{E_.^* - \frac{\tau}{T} E_. }{ 1- \frac{\tau}{T}}  \right)  } 
\label{ptauEs}
\end{eqnarray}
Plugging the Taylor expansion at first order in the ratio $\frac{\tau}{T} \ll 1$
\begin{eqnarray}
 S^{[2.5]}  \left( \frac{E_.^* - \frac{\tau}{T} E_. }{ 1- \frac{\tau}{T}}  \right)  
&&  =  S^{[2.5]}  \left( E_.^* + \frac{\tau}{T} (E_.^*-E_. ) + o \left( \frac{\tau^2}{T^2} \right)  \right)  
\nonumber \\
&&  = S^{[2.5]}(E_.^*) + \frac{\tau}{T} \sum_n (E_n^*-E_n ) \frac{ \partial S^{[2.5]}(E_.^*)}{\partial E_n^* }+ o \left( \frac{\tau^2}{T^2} \right)
\label{taylor}
\end{eqnarray}
into Eq. \ref{ptauEs} yields
\begin{eqnarray}
p_{\tau}(E_.) && \simeq C^{[2.5]}(E_.) e^{\tau S^{[2.5]}(E_.) }  
 e^{- T S^{[2.5]}(E_.^*) + (T-\tau) \left[ S^{[2.5]}(E_.^*) 
 + \frac{\tau}{T} \sum_n (E_n^*-E_n ) \frac{ \partial S^{[2.5]}(E_.^*)}{\partial E_n^* }
 + o \left( \frac{\tau^2}{T^2} \right)\right]   } 
 \nonumber \\
 && \opsimeq_{T \to +\infty}  C^{[2.5]}(E_.) e^{\tau \left[ S^{[2.5]}(E_.) - S^{[2.5]}(E_.^*)
 - \sum_n (E_n-E_n^* )  \frac{ \partial S^{[2.5]}(E_.^*)}{\partial E_n^* } \right]   }  
\label{ptauEfinal}
\end{eqnarray}
The factor of $\tau$ in the exponential
can be simplified via the introduction of the Markov model $M^{E^*_.}$ that makes 
the empirical observables $E^*_n$ typical (Eq. \ref{modeleeff})
\begin{eqnarray}
   E_n^* = E_n^{typ[M^{E^*_.}]}
\label{modeleeffstar}
\end{eqnarray}
in order to use the expression of the entropy of Eq. \ref{entropynonlinear}
\begin{eqnarray}
 S^{[2.5]}( E_.^*)= \sum_n E_n^* i_n(M^{E^*_.})
\label{entropynonlinearstar}
\end{eqnarray}
and of its derivatives with respect to empirical observables of Eq. \ref{derientropyempirical}
\begin{eqnarray}
\frac{ \partial  S^{[2.5]} \left(E^*_. \right) }{\partial E^*_n} = i_n(M^{E^*_.})  
\label{derientropyempiricalstar}
\end{eqnarray}
Plugging these two equations into Eq. \ref{ptauEfinal} leads to the final result
\begin{eqnarray}
p_{\tau}(E_.) 
 && \opsimeq_{T \to +\infty}  C^{[2.5]}(E_.) e^{\tau \left[ S^{[2.5]}(E_.) - \sum_n E_n^* i_n(M^{E^*_.})
 - \sum_n (E_n-E_n^* )  i_n(M^{E^*_.})  \right]   }  
 \nonumber \\
 && \opsimeq_{T \to +\infty}  C^{[2.5]}(E_.) e^{\tau \left[ S^{[2.5]}(E_.)   - \sum_n E_n  i_n(M^{E^*_.})  \right]   }  
 = C^{[2.5]}(E_.) \ e^{  - \tau  I^{[2.5]}_{M^{E^*_.}}  (E_.)    }
\label{ptauEfinalcano}
\end{eqnarray}
where one recognizes the Level 2.5 of Eqs \ref{level2.5} and \ref{rate2.5}
for the distribution of the empirical observables $E_n$ 
in the Canonical Ensemble associated to the Markov generator $M^{E^*_.} $ over the time-window $\tau$.

In conclusion, when the total trajectory $x(0 \leq t \leq T) $ belongs to
 the Microcanonical Ensemble of Eq. \ref{ptrajmicrot} corresponding to the fixed 
empirical observables $E_n^*$, 
the statistical properties of the subtrajectory $x(0 \leq t \leq \tau) $
over the time-window $\tau$ satisfying $1 \ll \tau \ll T$
are governed by the Canonical Ensemble associated to the Markov generator $M^{E^*_.} $


\subsubsection{  Time-additive observables in the Microcanonical Ensemble at Level 2.5}

In the Canonical Ensemble associated to the Markov generator $M$, 
the additive observables defined by Eq. \ref{additive} 
\begin{eqnarray}
A_M \left( E_.\right) = \sum_n E_n  a_n(M)
\label{additivecano}
\end{eqnarray}
are fluctuating with large deviations properties governed by Eq. \ref{additiveproba}.

The Markov model $M^{E^*_.}$ that makes 
the empirical observables $E^*_n$ typical (Eq. \ref{modeleeffstar})
is thus also useful to translate the definition of Eq. \ref{additivecano}
in the Canonical Ensemble 
in order to
obtain their fixed values in the Microcanonical Ensemble associated to the fixed empirical observables $E_n^*$
\begin{eqnarray}
A^{Micro[2.5]}_{E^*} \left( E_.\right) = \sum_n E^*_n  a_n(M^{E^*_.})
\label{additivemicro}
\end{eqnarray}


\subsection{ Maximization of the explicit entropy $ S^{[2.5]}( E_.) $ at Level 2.5 with constraints towards lower levels }

\label{subsec_maxi}

For the Canonical Ensemble, we have recalled in subsection \ref{subsec_contractionLD}
how the explicit rate function $ I_M^{[2.5]}( E_.) $ at Level 2.5 could be contracted
to obtain the large deviations properties of all the lower levels.
For the Microcanonical Ensemble, the analog idea 
concerns the maximization of the entropy $ S^{[2.5]}( E_.) $ at Level 2.5 with constraints
in order to obtain the entropies of all the lower levels with the following physical meaning.

From Eq. \ref{omegat} giving the explicit number $ \Omega^{[2.5]}_T( E_.) $ of trajectories of duration $T$ 
associated to given values of all the relevant empirical observables $E_n$,
one can compute :

(i) the number of trajectories of duration $T$
associated with any subset of the relevant empirical observables $E_n$
via the integration of $ \Omega^{[2.5]}_T( E_.) $
over all the empirical observables that one does not wish to keep.
For instance, the number $ \Omega^{[2]}_T( \rho_.) $
of trajectories of duration $T$ associated to the given empirical density $\rho_.$
can be obtained via the integration of $\Omega^{[2.5]}_T( E_.) $ over all the empirical flows
with their constraints
to obtain the Level 2
\begin{eqnarray}
\Omega^{[2]}_T  (\rho_.) 
\equiv \sum_{x(0 \leq t \leq T)}  \left[ \prod_x \delta \left( \rho_x - \rho_x \left[ x(.) \right] \right) \right] 
\opsimeq_{T \to +\infty}   C^{[2]}(\rho_.) \ e^{\displaystyle   T  S^{[2]}  (\rho_.)    }
\label{omegalevel2}
\end{eqnarray}
where the constitutive constraint $C^{[2]}(\rho_.)$ 
 corresponds to the normalization
of the empirical density $\rho_.$ as in Eq. \ref{level2},
while $S^{[2]}  (\rho_.)  $ represents the entropy as a function of the empirical density $\rho_.$ alone.

(ii) the number $\Omega^{[Add]}_T(A) $ of trajectories of duration $T$ corresponding to a given value $A$
of the additive observable of Eq. \ref{additivemicro}
can be obtained from the integration of $ \Omega^{[2.5]}_T( E_.) $ with the appropriate constraint
fixing the value of $A$
\begin{eqnarray}
\Omega^{[Add]}_T(A)  
&& = \sum_{ E_. }  \Omega^{[2.5]}_T( E_.)  \delta \left(A - \sum_n E_n  a_n(M^{E_.}) \right)
\nonumber \\
&& \opsimeq_{T \to +\infty}  \sum_{ E_. }  C^{[2.5]}(E_.) \ e^{\displaystyle   T  S^{[2.5]}  (E_.)    }
 \delta \left(A - \sum_n E_n  a_n(M^{E_.}) \right)
 \opsimeq_{T \to +\infty}  e^{ T S^{[Add]}(A) }
\label{additiveomega}
\end{eqnarray}
The entropy $S^{[Add]}(A)  $ thus
corresponds to the maximization of the entropy $S^{[2.5]} (E_.)  $
at Level 2.5
over the empirical observables $E_n$ satisfying the constitutive constraints $C^{[2.5]}(E_.) $
and the constraint $A = \sum_n E_n  a_n(M^{E_.}) $ reproducing the correct value $A $ of the additive observable
\begin{eqnarray}
S^{[Add]}(A)   = \opmax_{   
\substack{E_. :  C^{[2.5]}(E_.) {\rm and } \\ \sum_n E_n  a_n(M^{E_.})=A}}
  \left[  S^{[2.5]}  (E_.)  \right]
\label{additiveentropyfrom2.5}
\end{eqnarray}

(iii)  the Level 1 concerns the additive observables $A$ that can be reconstructed from
the empirical density $\rho_.$ alone, so that their number $ \Omega^{[1]}_T(A)  $
can be obtained from the integration of the Level 2 of Eq. \ref{omegalevel2}.

(iii) Finally, the Level 0 concerns 
  the total number of trajectories of duration $T$ when one integrates $ \Omega^{[2.5]}_T( E_.) $  over all thee empirical observables $E_.$
  or when one integrates $\Omega^{[2]}_T  (\rho_.)  $ over all possible empirical densities $\rho_.$,
  or when one integrates $\Omega^{[Add]}_T  (A)  $ over all possible values $A$
\begin{eqnarray}
\Omega^{[0]}_T \equiv  \sum_{ E_. }  C^{[2.5]}(E_.) \ e^{\displaystyle   T  S^{[2.5]}  (E_.)    } \opsimeq_{T \to +\infty}    e^{\displaystyle   T  S^{[0]}   }
\label{omegalevel0}
\end{eqnarray}
The entropy $S^{[0]}$ already appears in Ruelle the thermodynamic formalism
as discussed with simple examples in \cite{c_ruelle}.


\section{ Application to continuous-time Markov jump processes }

\label{sec_jump}

\subsection{ Canonical Ensemble of trajectories $x(0 \leq t \leq T)$ associated to the Markov jump generator $w$   }

\subsubsection{  Markov jump process converging towards some normalizable steady state  }

\label{subsec_jumpgenerator}

The continuous-time jump dynamics is defined by the Master Equation in discrete configuration space
\begin{eqnarray}
\partial_t P_x(t) =    \sum_{y }   w_{x,y}  P_y(t) 
\label{mastereq}
\end{eqnarray}
where the off-diagonal $x \ne y$ positive matrix elements $w_{x,y} \geq 0 $  represent the jump rates 
per unit time from $y$ towards $x \ne y$,
while the corresponding diagonal elements are negative and are fixed by the conservation of probability to be
\begin{eqnarray}
w_{y,y}  && =  - \sum_{x \ne y} w_{x,y} 
\label{wdiag}
\end{eqnarray}
The steady-state $P^*_x$ of Eq. \ref{mastereq}
\begin{eqnarray}
0 =    \sum_{y }   w_{x,y}  P_y^* 
\label{mastereqst}
\end{eqnarray}
 is assumed to be normalizable
\begin{eqnarray}
1 =    \sum_{y }  P_y^* 
\label{normast}
\end{eqnarray}

Here the elementary jump from $y$ towards $x$ is possible
 if the corresponding off-diagonal element is positive $w_{x,y}>0$,
 while it is impossible if $w_{x,y}=0$.
In particular, it is important to mention the following special cases :

(i) Directed dynamics : for any pair $x \ne y$, 
the two elementary jumps between them are never both possible 
\begin{eqnarray}
{\rm Directed: } \ \ \ w_{x,y} w_{y,x}=0
\label{directedw}
\end{eqnarray}

(ii) Undirected dynamics : for any pair $x \ne y$, 
the two jumps between them are either both possible $w_{x,y} w_{y,x}>0$
or both impossible $w_{x,y} =w_{y,x}=0$
\begin{eqnarray}
{\rm Undirected: } \ \ \ w_{x,y} w_{y,x}>0 \ \ \ {\rm or } \ \ w_{x,y} =w_{y,x}=0
\label{reversiblew}
\end{eqnarray}

(iii) Equilibrium dynamics : special case of undirected dynamics where 
on each link $x \ne y$ with possible moves $w_{x,y} w_{y,x}>0$,
there is no steady current in the steady state,
i.e. the two steady flows on each link satisfy the Detailed-Balance condition
\begin{eqnarray}
{\rm Detailed \ Balance: } \ \ \ w_{x,y} P_y^* = w_{y,x} P_x^*
\label{DetailedBalancew}
\end{eqnarray}


\subsubsection{Trajectories probabilities and their normalization} 

A trajectory $x(t)$ on the time interval $0 \leq t \leq T$
can be decomposed into a certain number $K \geq 0 $ of jumps $k=1,..,K$ occurring at times $0<t_1<...<t_K<T$
between the successive configurations $(x_0 \to x_1 \to x_2.. \to x_K)$ that are visited between these jumps.
The probability density of this trajectory for fixed initial condition $x_0$
\begin{eqnarray}
x(0 \leq t \leq T) = \left( x_0; t_1 ; x_1; t_2 ;... ; x_{K-1} ; t_K ; x_K \right)
\label{traject}
\end{eqnarray}
reads in terms of the matrix elements of the Markov generator $w_{.,.}$
\begin{eqnarray}
&& {\cal P}[x(0 \leq t \leq T)=\left( x_0; t_1 ; x_1; t_2 ;... ; x_{K-1} ; t_{K} ; x_K \right) ]
\nonumber \\
&& = 
 e^{ (T-t_K) w_{x_K,x_K} } 
 w_{x_K , x_{K-1} } 
 e^{ (t_K-t_{K-1} ) w_{x_{K-1},x_{K-1}} } 
 ...
  .... w_{x^{[2]} , x_{1} } 
 e^{ (t_2-t_1) w_{x_1,x_1} } 
 w_{x_1 ,x_{0} }
 e^{ t_1 w_{x_0,x_0} }  
 \nonumber \\
&& =  e^{ (T-t_K) w_{x_K,x_K} } \prod_{k=1}^K \left[ w_{x_k , x_{k-1} } e^{ (t_k-t_{k-1} ) w_{x_{k-1},x_{k-1}} } \right]
\label{ptraject}
\end{eqnarray}

The normalization over all possibles trajectories on $[0,T]$
involves the sum over the number $K$ of jumps, the sum 
over the $K$ configurations $(x_1,...,x_K$)
where $x_k$ has to be different from $x_{k-1}$, 
and the integration over the jump times with the measure $dt_1... dt_K$ and the constraint $0<t_1<...<t_K<T $
\begin{eqnarray}
 1  = && \sum_{K=0}^{+\infty}  \int_0^T dt_K \int_0^{t_K} dt_{K-1} ... \int_0^{t_2} dt_{1} 
 \sum_{x_K \ne x_{K-1}}\sum_{x_{K-1} \ne x_{K-2}}  
...
\sum_{x_2 \ne x_1}  \sum_{x_1 \ne x_0} 
 \nonumber \\ && 
 {\cal P}[x(0 \leq t \leq T)=\left( x_0; t_1 ; x_1; t_2 ;... ; x_{K-1} ; t_{K} ; x_K \right) ]
\label{norma}
\end{eqnarray}


\subsubsection{Identification of the relevant empirical observables that determine the trajectories probabilities }

The trajectory probability of Eq. \ref{ptraject}
 can be rewritten more compactly without the explicit enumeration of all the jumps as
\begin{eqnarray}
{\cal P }_w\left[ x(0 \leq t \leq T) \right]  
=  e^{ \displaystyle  \left[  \sum_{t \in [0,T] : x(t^+) \ne x(t^-) } \ln ( w_{x(t^+) , x(t^-) } ) +  \int_0^T dt  w_{x(t) , x(t) }   \right]  } 
\label{pwtrajjump}
\end{eqnarray}
The corresponding information per unit time ${\cal I}_w\left[ x(0 \leq t \leq T) \right] $
 of Eq. \ref{informationdef}
\begin{eqnarray}
{\cal I}_w\left[ x(.) \right]  \equiv - \frac{ \ln \left( {\cal P}_{[w_{.,.}]}\left[ x(.) \right] \right) }{T}
=  - \frac{1}{T} \sum_{t \in [0,T] : x(t^+) \ne x(t^-) } \ln ( w_{x(t^+) , x(t^-) } ) 
 - \frac{1}{T}  \int_0^T dt  w_{x(t) , x(t) }  
\label{informationjump}
\end{eqnarray}
involves only the following empirical time-averaged observables :

(a) the empirical density
\begin{eqnarray}
 \rho_{x}   \equiv \frac{1}{T} \int_0^T dt \  \delta_{x(t),x}  
 \label{rho1pj}
\end{eqnarray}
measures the fraction of the time spent by the trajectory $x(0 \leq t \leq T) $ in each configuration $x$ 
and is normalized to unity
\begin{eqnarray}
\sum_x \rho_{x}    = 1
\label{rho1ptnormaj}
\end{eqnarray}
Its typical value is the steady state $P^*$ of Eq. \ref{mastereqst}
\begin{eqnarray}
 \rho_{x}^{typ[w]} = P_x^*
\label{rhotypjump}
\end{eqnarray}

(b) the empirical flows 
\begin{eqnarray}
q_{x,y} \equiv  \frac{1}{T} \sum_{t \in [0,T] : x(t^+) \ne x(t^-)} \delta_{x(t^+),x} \delta_{x(t^-),y} 
\label{jumpempiricaldensity}
\end{eqnarray}
represent the density of jumps
from $y$ towards $x \ne y$ seen in the trajectory $x(0 \leq t \leq T) $.
For large $T$, these empirical flows satisfy the following stationarity constraints : 
for any configuration $x$, the total incoming flow into $x$ is equal to the total outgoing flow out of $x$
(up to boundary terms of order $1/T$ that involve only the initial configuration at $t=0$ and the final configuration at time $t=T$)
\begin{eqnarray}
\sum_{y \ne x} q_{x,y}= \sum_{y \ne x} q_{y,x} 
\label{contrainteq}
\end{eqnarray}
The typical values of the empirical flows are the steady state flows of the master Eq. \ref{mastereqst}
\begin{eqnarray}
 q_{x,y}^{typ[w]} = w_{x,y} P_y^* = w_{x,y}  \rho_{x}^{typ[w]} 
\label{qtypjump}
\end{eqnarray}

With respect to the general formalism of section \ref{sec_general},
this means that 

(i) the relevant empirical observables $E_.$ 
that determine the trajectories probabilities
are the empirical density $\rho_.$ and the empirical flows $q_{.,.}$
\begin{eqnarray}
 E_. = \left( \rho_. ; q_{.,.} \right)
\label{Ejump}
\end{eqnarray}

(ii) their constitutive constraints $C^{[2.5]}(E_.)$ are given by Eqs \ref{rho1ptnormaj} and \ref{contrainteq}
\begin{eqnarray}
 C^{[2.5]}\left( \rho_. ; q_{.,.} \right)
  = \delta \left( \sum_x \rho_{x} - 1 \right) 
 \prod_x \delta \left( \sum_{y \ne x} q_{x,y} - \sum_{y \ne x} q_{y,x}  \right)
\label{constraints2.5master}
\end{eqnarray}

(iii) Eq. \ref{informationjump}
yields that the information $I_w(E_.)$ is given by the following linear combination
of the empirical observables $E_.=\left( \rho_.  ; q_{.,.} \right)$  
\begin{eqnarray}
I_w \left( \rho_.  ; q_{.,.} \right)
=       \sum_{y  }\sum_{ x \ne y  } \bigg( w_{x , y }  \rho_{y}   - q_{x,y}\ln ( w_{x,y } )      \bigg)
\equiv \sum_{y  } \rho_{y} i_y(w) + \sum_{y  }\sum_{ x \ne y  } q_{x,y}i_{x,y}(w) 
\label{informationempimaster}
\end{eqnarray}
The corresponding coefficients  that represent their intensive conjugate parameters
\begin{eqnarray}
i_y (w) && \equiv \sum_{ x \ne y  }  w_{x , y } = - w_{y,y}
\nonumber \\
i_{x,y}(w) && \equiv -  \ln \left(w_{x ,y} \right)
\label{lambdajump}
\end{eqnarray}
are very simple in terms of the matrix elements $w_{.,.}$ of the Markov generator.


\subsubsection{ Boltzmann intensive entropy $S^{[2.5]}\left( \rho_. ; q_{.,.} \right)$ as a function of the empirical observables $E_.=\left( \rho_.  ; q_{.,.} \right)$ }

Eq. \ref{qtypjump} yields that the modified Markov generator $w^E$ that would make
the empirical observables $E_.=\left( \rho_.  ; q_{.,.} \right)$ typical 
is given by the jump rates for $x \ne y$
\begin{eqnarray}
w^E_{x ,y} \equiv \frac{q_{x,y}}{ \rho_{y}} 
\label{wjumpinfer}
\end{eqnarray}
As a consequence,
Eq. \ref{compensation} yields the 
entropy $S^{[2.5]}\left( \rho_. ; q_{.,.} \right)$ as a function of the empirical observables $E_.=\left( \rho_.  ; q_{.,.} \right)$ reads using Eqs \ref{informationempimaster} and \ref{wjumpinfer}
\begin{eqnarray}
S^{[2.5]}\left( \rho_. ; q_{.,.} \right)    =  I_{w^E}  \left( \rho_. ; q_{.,.} \right)   
  =      \sum_{y  }\sum_{ x \ne y  } \left[ q_{x,y}  - q_{x,y}\ln \left( \frac{q_{x,y}}{ \rho_y}  \right)     \right]
\label{entropyempijump}
\end{eqnarray}
As stressed after Eq. \ref{entropynonlinear}, it is nonlinear with respect 
to the empirical observables $E_.=\left( \rho_.  ; q_{.,.} \right)$.


\subsubsection{ Rate function at level 2.5 for the empirical observables $E_.=\left( \rho_.  ; q_{.,.} \right)$ }

The joint probability distribution of the empirical density $\rho_.$ and the empirical flows $q_{.,.}$
follows the large deviation form \cite{fortelle_thesis,fortelle_jump,maes_canonical,maes_onandbeyond,wynants_thesis,chetrite_formal,BFG1,BFG2,verley2016,chetrite_HDR,c_ring,c_interactions,chemical,c_open,c_detailed,barato_periodic,chetrite_periodic,c_reset,c_inference,c_runandtumble,c_jumpdiff,c_skew,c_metastable,c_east,c_exclusion}
\begin{eqnarray}
P^{[2.5]}_w \left( \rho_. ; q_{.,.} \right) \opsimeq_{T \to +\infty} 
C^{[2.5]}( \rho_. ; q_{.,.} ) e^{- T I^{[2.5]}_w( \rho_. ; q_{.,.} ) }
\label{level2.5master}
\end{eqnarray}
where the constitutive constraints $C^{[2.5]}( \rho_. ; q_{.,.} )$ have been written in Eq. \ref{constraints2.5master},
while the rate function at Level 2.5 is given by the difference between the information 
of Eq. \ref{informationempimaster}
and the entropy from Eq. \ref{entropyempijump}
\begin{eqnarray}
I^{[2.5]}_w( \rho_. ; q_{.,.} )&& = I_w \left(  \rho_. ; q_{.,.}  \right)  - S^{[2.5]} \left(  \rho_. ; q_{.,.}  \right) 
\nonumber \\
&& = \sum_{y } \sum_{x \ne y} 
\left[ q_{x,y}  \ln \left( \frac{ q_{x,y}  }{  w_{x,y}  \rho_y }  \right) 
 - q_{x,y}  + w_{x,y}  \rho_y  \right]
\label{rate2.5master}
\end{eqnarray}


\subsubsection{ Kolmogorov-Sinai entropy $h_w^{KS}$ }

The Kolmogorov-Sinai entropy $h^{KS}_w$ of Eqs \ref{hksempityp} and \ref{hksempitypSalso}
reads using the typical values of Eqs \ref{rhotypjump} and \ref{qtypjump}
\begin{eqnarray}
h^{KS}_w  =     \sum_{y  } P^*_{y} \sum_{ x \ne y  } w_{x , y }  \bigg(   1  - \ln ( w_{x,y } )      \bigg)
\label{hksempitypjump}
\end{eqnarray}
so that it can be computed whenever the steady state $P^*(y)$ associated to the generator $w$ is known (see \cite{c_ruelle} for more discussions with examples).


\subsection{ Microcanonical Ensemble at Level 2.5 with fixed empirical density and flows }

\subsubsection{ Microcanonical Ensemble at Level 2.5 where the empirical density and flows cannot fluctuate for finite $T$}

In the Microcanonical Ensemble of Eq. \ref{microcanodeltaempirical}
 \begin{eqnarray}
P^{Micro[2.5]}_{\left( \rho^*_. ; q^*_{.,.} \right)} \left( \rho_. ; q_{.,.} \right)    
 = \left[ \prod_x \delta \left( \rho_x - \rho_x^* \right) \right]
 \left[ \prod_y  \prod_{x \ne y} \delta \left( q_{x,y} - q_{x,y}^* \right) \right]
\label{microempijump}
\end{eqnarray}
the empirical observables $\left( \rho_. ; q_{.,.} \right)$ are fixed to given values
$\left( \rho^{*}_. ; q^{*}_{.,.} \right) $ satisfying the constitutive constraints of Eq. \ref{constraints2.5master}
\begin{eqnarray}
 C^{[2.5]}\left( \rho^*_. ; q^*_{.,.} \right)
  = \delta \left( \sum_x \rho^*_{x} - 1 \right) 
 \prod_x \delta \left( \sum_{y \ne x} q^*_{x,y} - \sum_{y \ne x} q^*_{y,x}  \right)
\label{constraints2.5etoile}
\end{eqnarray}
and cannot fluctuate for finite $T$, in contrast to Eq. \ref{level2.5master} in the Canonical Ensemble associated to the Markov generator $w$.

Here the jump from $y$ towards $x\ne y$ is possible
 if the corresponding fixed empirical flow is positive $q^*_{x,y}>0$,
 while it is impossible if $q^*_{x,y}=0$.

In particular, the two special cases of Eqs \ref{directedw} and \ref{reversiblew}  
mentioned at the end of subsection \ref{subsec_jumpgenerator}
translate into :

(i) Directed dynamics : for any pair $x \ne y$, 
the two jumps between them are never both possible 
\begin{eqnarray}
{\rm Directed: } \ \ \ q^*_{x,y} q^*_{y,x}=0
\label{directedq}
\end{eqnarray}

(ii) Undirected dynamics : for any pair $x \ne y$, 
the two jumps between them are either both possible $q^*_{x,y} q^*_{y,x}>0$
or both impossible $q^*_{x,y} =q^*_{y,x}=0$
\begin{eqnarray}
{\rm Undirected: } \ \ \ q^*_{x,y} q^*_{y,x}>0 \ \ \ {\rm or } \ \ q^*_{x,y} =q^*_{y,x}=0
\label{reversibleq}
\end{eqnarray}

(iii) Equilibrium dynamics : special case of undirected dynamics where 
on each link $x \ne y$ with possible flows $q^*_{x,y} q^*_{y,x}>0$,
 these two empirical flows on each link satisfy the Detailed-Balance condition
\begin{eqnarray}
{\rm Detailed \ Balance: } \ \ \ q^*_{x,y} = q^*_{y,x}
\label{DetailedBalanceq}
\end{eqnarray}


\subsubsection{ Probabilities of trajectories $x(0 \leq t \leq T) $ in the Microcanonical Ensemble  }

In the Microcanonical Ensemble of Eq. \ref{ptrajmicro}, only the trajectories $[x(0 \leq t \leq T)]  $
corresponding to the empirical values $\left( \rho^{*}_. ; q^{*}_{.,.} \right) $ have a non-zero weight,
and all these allowed trajectories have the same weight given by the inverse of their number $\Omega^{[2.5]}_T\left( \rho^*_. ; q^*_{.,.} \right) $ of Eq. \ref{omegaempi}
\begin{eqnarray}
{\cal P}^{Micro[2.5]}_{\left( \rho^*_. ; q^*_{.,.} \right)}[x(.)]   
 =  \frac{1 }{\Omega^{[2.5]}_T\left( \rho^*_. ; q^*_{.,.} \right) }
 \left[ \prod_x \delta \left( \rho^*_x - \rho_x\left[ x(.) \right] \right) \right]
 \left[ \prod_y  \prod_{x \ne y} \delta \left( q^*_{x,y} - q_{x,y}\left[ x(.) \right] \right) \right]
\label{ptrajmicrojump}
\end{eqnarray}
For large $T$, Eq. \ref{ptrajmicro} reads
\begin{eqnarray}
{\cal P}^{Micro[2.5]}_{\left( \rho^*_. ; q^*_{.,.} \right)}[x(.)]   
&& \opsimeq_{T \to +\infty} 
 e^{\displaystyle - T \ S^{[2.5]}\left( \rho^*_. ; q^*_{.,.} \right)  }
\left[ \prod_x \delta \left( \rho^*_x - \rho_x\left[ x(.) \right] \right) \right]
 \left[ \prod_y  \prod_{x \ne y} \delta \left( q^*_{x,y} - q_{x,y}\left[ x(.) \right] \right) \right]
\label{ptrajmicrotjump}
\end{eqnarray}
with the entropy $S^{[2.5]}\left( \rho^*_. ; q^*_{.,.} \right) $ of Eq. \ref{entropyempijump}
\begin{eqnarray}
S^{[2.5]} \left( \rho^*_. ; q^*_{.,.} \right)  
  =      \sum_{y  }\sum_{ x \ne y  } \left[ q^*_{x,y}  - q^*_{x,y}\ln \left( \frac{q^*_{x,y}}{ \rho^*_{y}}  \right)     \right]
\label{entropyempistar}
\end{eqnarray}


\subsubsection{ Statistics of the subtrajectories on $[0,\tau]$ of the Microcanonical Ensemble trajectories on $[0,T]$ for $1 \ll \tau \ll T$ }

As explained in details in subsection \ref{subsec_subtraj},
the fact that the Canonical Ensemble emerges to describe the statistics
of the subtrajectories is based on the property of Eq. \ref{derientropyempirical} 
concerning 
the derivatives of the entropy $S^{[2.5]}\left(E_. \right) $ with respect to the empirical observables $E_n$.
For Markov jump processes, the derivatives of the entropy $S^{[2.5]}\left( \rho_. ; q_{.,.} \right) $ of Eq. \ref{entropyempijump}
with respect to the empirical density $\rho_y$ and with respect to the empirical flows $q_{x,y}$
can be indeed rewritten in terms of the modified generator $w^E$ of Eq. \ref{wjumpinfer}
and correspond to the coefficients of Eq. \ref{lambdajump}
\begin{eqnarray}
\frac{ \partial S^{[2.5]}\left( \rho_. ; q_{.,.} \right)  }{\partial \rho_y}    
&&  =     \frac{1}{\rho_y} \sum_{ x \ne y  }  q_{x,y}  
  = \sum_{ x \ne y  } w^E_{x ,y} = - w^E_{y ,y} = i_y(w^E)
  \nonumber \\
\frac{ \partial S^{[2.5]}\left( \rho_. ; q_{.,.} \right)  }{\partial q_{x,y}}   
&& =    - \ln \left( \frac{q_{x,y}}{ \rho_y}  \right)   
= - \ln \left( w^E_{x ,y}  \right)   
  = i_{x,y}(w^E)
  \label{entropyempijumpderi}
\end{eqnarray}
in agreement with the general property of Eq. \ref{derientropyempirical}.


\subsection{Simple example of continuous-time directed trajectories on a ring : entropies at various levels  }

The simplest example of continuous-time directed trajectories is based on
a ring of $L$ sites with periodic boundary conditions $x+L \equiv x$ :
the empirical flow $q_{x,y}$ is non-vanishing only for $x=y+1$
\begin{eqnarray}
q_{x,y} =  \delta_{x,y+1} q_{y+1,y}
\label{qtrap}
\end{eqnarray}
and the stationarity constraints of Eq. \ref{contrainteq}
yields that the $L$ elements $ q_{x+1,x} $
take the same value $j$ along the ring $x=1,2,..,L$, that represents the current $j$
flowing through each link of the ring
\begin{eqnarray}
  q_{x+1,x} = j
\label{qjtrap}
\end{eqnarray}

\subsubsection{ Number $\Omega^{[2.5]}_T\left( \rho_{.} ,j\right) $ of trajectories with the empirical density $\rho_.$ and the current $j$ } 

Since the empirical flows $q_{.,.}$ of Eqs \ref{qtrap} and \ref{qjtrap} involve a single positive parameter $j \in [0,+\infty[$,
the entropy of Eq. \ref{entropyempijump} reduces to
\begin{eqnarray}
S^{[2.5]} \left( \rho_. ; j \right)  
  =      \sum_{y=1  }^L \left[ j - j \ln \left( \frac{j}{ \rho_{y}}  \right)     \right]
  =  L \left[ j - j \ln(j) \right] + j  \sum_{y=1  }^L \ln(\rho_y)
\label{entropy2.5jumptrap}
\end{eqnarray}
and
the number $\Omega^{[2.5]}_T\left( \rho_{.} ,j\right) $ of trajectories  
 reads
\begin{eqnarray}
\Omega^{[2.5]}_T\left( \rho_{.} ,j\right)
 \opsimeq_{T \to +\infty}  \delta \left( \sum_{x=1}^L \rho_x -1\right) \theta(j)
 e^{\displaystyle  T  S^{[2.5]} \left( \rho_{.} ,j \right)  }
 \label{trapomega2.5jump}
\end{eqnarray}

\subsubsection{ Number $\Omega^{[2]}_T\left( \rho_{.} \right) $ of trajectories with the empirical density $\rho_.$  } 

The number of trajectories $\Omega^{[2]}_T\left( \rho_{.} \right) $ 
with the empirical density $\rho_.$ can be computed via the integration 
of Eq. \ref{trapomega2.5jump} over the current $j$
\begin{eqnarray}
\Omega^{[2]}_T\left( \rho_{.} \right) && = \int dj \Omega^{[2.5]}_T\left( \rho_{.} ,j\right)
 \opsimeq_{T \to +\infty} 
   \delta \left( \sum_{x=1}^L \rho_x -1\right) 
   \int_0^{+\infty} dj  
 e^{\displaystyle  T  S^{[2.5]} \left( \rho_{.} ,j \right)  }
\nonumber \\
&&   \opsimeq_{T \to +\infty} 
   \delta \left( \sum_{x=1}^L \rho_x -1\right) 
   e^{\displaystyle  T  S^{[2]} \left( \rho_{.}  \right)  }
 \label{trapomega2jump}
\end{eqnarray}
The optimization of the entropy $S^{[2.5]} \left( \rho_{.} ,j \right) $ of Eq. \ref{entropy2.5jumptrap}
 over the current $j$
\begin{eqnarray}
0 = \frac{ \partial S^{[2.5]} \left( \rho_{.} ,j \right)  }{ \partial j}
  =   -  L  \ln(j)  +   \sum_{y=1  }^L \ln(\rho_y)
\label{entropy2.5trapderiJjump}
\end{eqnarray}
leads to the optimal current
\begin{eqnarray}
j^{opt} = e^{ \displaystyle  \frac{1}{L}  \sum_{y=1  }^L \ln(\rho_y)} = \left[ \prod_{y=1}^L \rho_y \right]^{\frac{1}{L} }
\label{trapjoptjump}
\end{eqnarray}
that can be plugged into Eq. \ref{entropy2.5jumptrap}
to obtain the entropy at Level 2 that governs Eq. \ref{trapomega2jump}
\begin{eqnarray}
S^{[2]} \left( \rho_{.}  \right) = S^{[2.5]} \left( \rho_{.} ,j^{opt} \right)  
  =  L  j^{opt} = L \left[ \prod_{y=1}^L \rho_y \right]^{\frac{1}{L} }
  \label{entropy2trapjump}
\end{eqnarray}

\subsubsection{ Number $\Omega^{[Curr]}_T\left( j\right) $ of trajectories 
with the empirical current $j$  } 

The number of trajectories $\Omega^{[Current]}_T\left( j\right) $ 
with the empirical current $j$  can be computed via the integration 
of Eq. \ref{trapomega2.5jump} over the empirical density $\rho_.$
\begin{eqnarray}
\Omega^{[Current]}_T\left( j\right)  && = \int d\rho_. \Omega^{[2.5]}_T\left( \rho_{.} ,j\right)
 \opsimeq_{T \to +\infty} \theta(j) \int d\rho_.
   \delta \left( \sum_{x=1}^L \rho_x -1\right) 
  e^{\displaystyle  T  S^{[2.5]} \left( \rho_{.} ,j \right)  }
\nonumber \\
&&   \opsimeq_{T \to +\infty} \theta(j) 
  e^{\displaystyle  T  S^{[Current]} \left( j \right)  }
 \label{trapomegajjump}
\end{eqnarray}
To optimize the entropy $S^{[2.5]} \left( \rho_{.} ,J \right) $ of Eq. \ref{entropy2.5jumptrap}
over the empirical density $\rho_.$
satisfying the normalization constraint, one introduces the following Lagrangian containing the Lagrange multiplier $\mu$
\begin{eqnarray}
{\cal L}_j(\rho_.) && \equiv S^{[2.5]} \left( \rho_{.} ,j \right)  -\mu \left( \sum_{y=1}^L \rho_y -1\right) 
\nonumber \\
 &&  = 
  L \left[ j - j \ln(j) \right] + j  \sum_{y=1  }^L \ln(\rho_y)
  -\mu \left( \sum_{y=1}^L \rho_y -1\right)
\label{traplagrangianjump}
\end{eqnarray}
The optimization over $\rho_y$
\begin{eqnarray}
0=\frac{ \partial {\cal L}_j(\rho_.)}{\partial \rho_y}     =  
 \frac{j}{\rho_y}   -\mu 
\label{traplagrangianderijump}
\end{eqnarray}
yields that the optimal density $\rho_y^{opt}= \frac{j}{\mu}$ 
does not depend on the position $y$, so that its value is actually fixed by the normalization constraint
\begin{eqnarray}
 \rho_y^{opt} = \frac{1}{L}
\label{traprhounif}
\end{eqnarray}
Plugging this uniform optimal solution into Eq. \ref{entropy2.5jumptrap}
leads to the entropy $S^{[Current]}\left( j\right) $ that governs Eq. \ref{trapomegajjump}
\begin{eqnarray}
S^{[Current]}\left( j\right)=S^{[2.5]} \left( \rho^{opt}_{.} ,j \right)  
 =  j L \left[ 1 -  \ln(jL)  \right] 
\label{entropyjtrapjump}
\end{eqnarray}

\subsubsection{ Total number $\Omega^{[0]}_T $ of trajectories   } 

The total number $\Omega^{[0]}_T $ of trajectories can be computed from the integration
of Eq. \ref{trapomegajjump} over the current $j$
\begin{eqnarray}
\Omega^{[0]}_T = \int dj \Omega^{[Current]}_T\left( j\right)  
 \opsimeq_{T \to +\infty} \int_0^{+\infty} dj
 e^{\displaystyle  T  S^{[Current]}\left( j\right)  } \opsimeq_{T \to +\infty}e^{\displaystyle  T  S^{[0]}  }
 \label{trapomega0}
\end{eqnarray}
The optimization of Eq. \ref{entropyjtrapjump}
over the current $j$
\begin{eqnarray}
0= \frac{ \partial S^{[Current]}\left( j\right) }{\partial j} 
 =  - L  \ln(jL) 
\label{entropyJtrapderijump}
\end{eqnarray}
yields the optimal value
\begin{eqnarray}
j^{opt} = \frac{1}{L}
\label{Joptzerojump}
\end{eqnarray}
that can be plugged into Eq. \ref{entropyjtrapjump}
to obtain the entropy at Level 0 
\begin{eqnarray}
S^{[0]} = S^{[Current]}\left( j^{opt}\right)  
 = L j^{opt}=1
\label{entropytrap0jump}
\end{eqnarray}
This simple value can be understood from the integration of the measure on the first line of Eq. \ref{norma},
where the sums over the positions disappear as a consequence of the one-dimensional directed character of the present model
\begin{eqnarray}
 \sum_{K=0}^{+\infty}  \int_0^T dt_K \int_0^{t_K} dt_{K-1} ... \int_0^{t_2} dt_{1} 
=\sum_{K=0}^{+\infty} \frac{1}{K!} \prod_{k=1}^K  \left[\int_0^T dt_k  \right]
= \sum_{K=0}^{+\infty} \frac{T^K}{K!} =e^T = e^{T S^{[0]} }
\label{normawtdir}
\end{eqnarray}
in agreement with the entropy found in Eq. \ref{entropytrap0jump}.


\section{Undirected Markov Jump Processes :  explicit entropies at various levels  }

\label{sec_EntropyJump}

In this section, we focus on the case of undirected jump processes satisfying Eq. \ref{reversibleq},
where the contraction of the entropy at Level 2.5
can be implemented to obtain explicit expressions for many entropies of lower levels. 

\subsection{ Replacing the empirical flows $q_{.,.}$ by the empirical currents $j_{.,.}$ and 
the empirical activities $a_{.,.}$ }

On each link $x \ne y$ where the two flows are possible $q_{x,y} q_{y,x}>0$,
it is convenient to choose an order $x>y$ and to introduce the corresponding neighborhood notations $N_{.}^{\pm}$
\begin{eqnarray}
{\rm Order \ on \ each \ link } \ \ q_{x,y} q_{y,x}>0:  \ \ \  x \in N^+_y \ \ {\rm and} \ \  y \in N^-_x  
\label{reversibleqneighbor}
\end{eqnarray}
while the total neighborhood $N_y$ of $y$ is the reunion of $N^+_y$ and $N^-_y$.

The two non-vanishing empirical flows $q_{x,y}$ and $q_{y,x}$ can be replaced
by their antisymmetric and symmetric parts called respectively 
the empirical current $j_{x,y}  $ and the empirical activity $a_{x,y} $
\begin{eqnarray}
j_{x,y} && =q_{x,y} -  q_{y,x}= - j_{y,x}
\nonumber \\
a_{x,y}  && = q_{x,y} +  q_{y,x}  =  a_{y,x}
\label{jafromq}
\end{eqnarray}
The constitutive constraints of Eq. \ref{constraints2.5master} 
do not contain the activities $a_{.,.}$
and are factorized 
\begin{eqnarray}
 C^{[2.5]}\left( \rho_. ; j_{.,.} \right)
  = C^{[2]}\left( \rho_.  \right)
 C^{statio}\left(  j_{.,.}  \right)
\label{constraints2.5masterj}
\end{eqnarray}
into the normalization constraint for the empirical density $\rho_.$
\begin{eqnarray}
 C^{[2]}\left( \rho_.  \right)
  = \delta \left( \sum_x \rho_{x} - 1 \right) 
\label{c2jump}
\end{eqnarray}
and into the stationarity constraints for the empirical currents $j_{.,.}$
\begin{eqnarray}
 C^{statio}\left(  j_{.,.}  \right)
  = \prod_x \delta \left( \sum_{y \in N_x} j_{x,y}    \right)
=  \prod_x \delta \left( \sum_{y \in N_x^+} j_{x,y} + \sum_{y \in N_x^-} j_{x,y}  \right)
\label{cstatiojump}
\end{eqnarray}


\subsection{ Number $\Omega^{[2.5]}_T (\rho_. ; j_{.,.} ; a_{.,.}  ) $ of trajectories 
$x(.)$ with 
the empirical density $\rho_.$, currents $j_{.,.}$ and activities $a_{.,.}$ }

Via the change of variables of Eq. \ref{jafromq},
the number $\Omega^{[2.5]}_T (\rho_. ; j_{.,.} ; a_{.,.}  ) $ of trajectories $x(0 \leq t \leq T)$ with 
the empirical density $\rho_.$, the empirical currents $j_{.,.}$ and the empirical activities $a_{.,.}$
is given by
\begin{eqnarray}
 \Omega^{[2.5]}_T( \rho_. ; j_{.,.} ; a_{.,.} ) \opsimeq_{T \to +\infty} 
 C^{[2]}\left( \rho_.  \right) C^{statio}\left(  j_{.,.}  \right)
  \ e^{\displaystyle T S^{[2.5]}( \rho_. ; j_{.,.} ; a_{.,.}  )  }
\label{omegatjumpja}
\end{eqnarray}
where the constraints have been written in Eqs \ref{c2jump} and \ref{cstatiojump},
while the entropy $S^{[2.5]}( \rho_. ; j_{.,.} ; a_{.,.}  ) $ at Level 2.5 of Eq. \ref{entropyempistar}
becomes using the choice of some ordering $x>y$ on each link (Eq. \ref{reversibleqneighbor})
\begin{eqnarray}
S^{[2.5]} \left( \rho_. ;  j_{.,.} ; a_{.,.} \right)  
&&  =      \sum_{y  }\sum_{ x \in N^+_y  } \left[ a_{x,y}  
  - \frac{a_{x,y} + j_{x,y}}{2}\ln \left( \frac{a_{x,y} + j_{x,y}}{ 2 \rho_y}  \right)  
   - \frac{a_{x,y} - j_{x,y}}{2}\ln \left( \frac{a_{x,y} - j_{x,y}}{ 2\rho_x}  \right)    
        \right]
        \nonumber \\
        &&   =      \sum_{y  }\sum_{ x \in N^+_y  } \left[ a_{x,y}  
   - \frac{a_{x,y} }{2}\ln \left( \frac{a_{x,y}^2 - j_{x,y}^2}{ 4 \rho_x \rho_y}  \right)    
   - \frac{ j_{x,y}}{2}\ln \left( \frac{ (a_{x,y} + j_{x,y}) \rho_x }{ (a_{x,y} - j_{x,y}) \rho_y}  \right)  
     \right]
\label{entropyempiexplicitaj}
\end{eqnarray}

In order to see more clearly how the entropy $S^{[2.5]} \left( \rho_. ;  j_{.,.} ; a_{.,.} \right)   $
depends on the empirical currents $j_{.,.}$,
let us rewrite Eq. \ref{entropyempiexplicitaj} as a sum
\begin{eqnarray}
S^{[2.5]} \left( \rho_. ;  j_{.,.} ; a_{.,.} \right)  
           =      \sum_{y  }\sum_{ x \in N^+_y  } 
           S_{[x,y]}^{[2.5]} \left( \rho_x ; \rho_y ; j_{x,y} ; a_{x,y} \right)
           \label{entropyempiexplicitajintolinks}
\end{eqnarray}
of the contributions associated to the links $[x,y]$
\begin{eqnarray}
S_{[x,y]}^{[2.5]} \left( \rho_x ; \rho_y ; j_{x,y} ; a_{x,y} \right)  
          \equiv a_{x,y}  
   - \frac{a_{x,y} }{2}\ln \left( \frac{a_{x,y}^2 - j_{x,y}^2}{ 4 \rho_x \rho_y}  \right)    
   - \frac{ j_{x,y}}{2}\ln \left( \frac{ (a_{x,y} + j_{x,y}) \rho_x }{ (a_{x,y} - j_{x,y}) \rho_y}  \right)  
\label{entropyempiexplicitajlink}
\end{eqnarray}
The contribution of the link $[x,y]$ can be decomposed
into its even and its odd parts with respect to the link current $j_{x,y}$
\begin{eqnarray}
S_{[x,y]}^{[2.5]} \left( \rho_x ; \rho_y ; j_{x,y} ; a_{x,y} \right)  
          =   S_{[x,y]even}^{[2.5]} \left( \rho_x ; \rho_y ; j_{x,y} ; a_{x,y} \right)  
          +S_{[x,y]odd}^{[2.5]} \left( \rho_x ; \rho_y ; j_{x,y} ; a_{x,y} \right)  
\label{entropyempiexplicitajlinkevenodd}
\end{eqnarray}
The even contribution reads
\begin{eqnarray}
   S_{[x,y]even}^{[2.5]} \left( \rho_x ; \rho_y ; j_{x,y} ; a_{x,y} \right)  
&& = \frac{S_{[x,y]}^{[2.5]} \left( \rho_x ; \rho_y ; j_{x,y} ; a_{x,y} \right) 
+ S_{[x,y]}^{[2.5]} \left( \rho_x ; \rho_y ; - j_{x,y} ; a_{x,y} \right)}{2} 
\nonumber \\
&& =   a_{x,y}  
   - \frac{a_{x,y} }{2}\ln \left( \frac{a_{x,y}^2 - j_{x,y}^2}{ 4 \rho_x \rho_y}  \right)    
   - \frac{ j_{x,y}}{2}\ln \left( \frac{ a_{x,y} + j_{x,y}  }{ a_{x,y} - j_{x,y} }  \right)  
  \label{entropyempiexplicitajlinkeven}
\end{eqnarray}
while the odd contribution
\begin{eqnarray}
   S_{[x,y]odd}^{[2.5]} \left( \rho_x ; \rho_y ; j_{x,y} ; a_{x,y} \right)  
&& = \frac{S_{[x,y]}^{[2.5]} \left( \rho_x ; \rho_y ; j_{x,y} ; a_{x,y} \right) 
- S_{[x,y]}^{[2.5]} \left( \rho_x ; \rho_y ; - j_{x,y} ; a_{x,y} \right)}{2} 
\nonumber \\
&&=  \frac{ j_{x,y}}{2}\ln \left( \frac{  \rho_y }{  \rho_x}  \right) 
\label{entropyempiexplicitajlinkodd}
\end{eqnarray}
is simply linear in the current $j_{x,y}$.
The sum of all these odd links contributions
vanishes as a consequence of the antisymmetry of the current $j_{x,y}=-j_{y,x}$
of Eq. \ref{jafromq} and of the stationarity constraint $C^{statio}(j_{.,.})$ of Eq. \ref{cstatiojump}
\begin{eqnarray}
  \sum_{y  }\sum_{ x \in N^+_y  }    S_{[x,y]odd}^{[2.5]} \left( \rho_x ; \rho_y ; j_{x,y} ; a_{x,y} \right)  
&& =   \sum_{y  }\sum_{ x \in N^+_y  }   \frac{ j_{x,y}}{2} \ln ( \rho_y )  
-   \sum_{y  }\sum_{ x \in N^+_y  }  \frac{ j_{x,y}}{2} \ln ( \rho_x )  
\nonumber \\
&& = -  \sum_{x  } \frac{\ln ( \rho_x ) }{2}  \left[ \sum_{ y \in N_x^+  }   j_{x,y}  
+   \sum_{ y \in N_x^-  }   j_{x,y} \right]
= -  \sum_{x  } \frac{\ln ( \rho_x ) }{2}  \left[ \sum_{ y \in N_x  }   j_{x,y}   \right] =0
\ \ \ 
\label{entropyempiexplicitajlinkoddsum}
\end{eqnarray}
As a consequence, the entropy of Eq. \ref{entropyempiexplicitajintolinks}
reduces to the sum of the even contributions of the links
\begin{eqnarray}
S^{[2.5]} \left( \rho_. ;  j_{.,.} ; a_{.,.} \right)  
  &&         =      \sum_{y  }\sum_{ x \in N^+_y  } 
           S_{[x,y]even}^{[2.5]} \left( \rho_x ; \rho_y ; j_{x,y} ; a_{x,y} \right)   
\nonumber \\
&&           =   \sum_{y  }\sum_{ x \in N^+_y  } 
\left[ a_{x,y}  
   - \frac{a_{x,y} }{2}\ln \left( \frac{a_{x,y}^2 - j_{x,y}^2}{ 4 \rho_x \rho_y}  \right)    
   - \frac{ j_{x,y}}{2}\ln \left( \frac{ a_{x,y} + j_{x,y}  }{ a_{x,y} - j_{x,y} }  \right)   \right]          
                    \label{entropyempiexplicitajintolinksevenonly}
\end{eqnarray}
The physical meaning is that once the empirical density $\rho_.$ and the empirical activity $a_{.,.}$ are given,
any realization of the empirical links currents $j_{x,y}$ that satisfies the stationary constraints of Eq. \ref{cstatiojump}
has the same entropy as the configuration with the reversed empirical link currents $(-j_{x,y})$ that also satisfies the stationary constraints 
\begin{eqnarray}
S^{[2.5]} \left( \rho_. ;  j_{.,.} ; a_{.,.} \right)  = S^{[2.5]} \left( \rho_. ;  -j_{.,.} ; a_{.,.} \right)  
\label{entropy2.5eveninj}
\end{eqnarray}


\subsection{ Number $\Omega^{[2.25]}_T (\rho_. ; a_{.,.}  ) $ of trajectories 
$x(.)$ with 
the empirical density $\rho_.$ and the empirical activities $a_{.,.}$   }

The number $\Omega^{[2.25]}_T (\rho_. ; a_{.,.}   ) $
of trajectories $x(0 \leq t \leq T)$ with given empirical density $\rho_.$ and given empirical activities $a_{.,.}$
can be obtained via the integration of Eq. \ref{omegatjumpja} over the currents $j_{.,.}$
\begin{eqnarray}
\Omega^{[2.25]}_T (\rho_. ; a_{.,.}   )&& \equiv 
 \int d j_{.,.} \Omega^{[2.5]}_T( \rho_. ; j_{.,.} ; a_{.,.} ) \opsimeq_{T \to +\infty} 
 C^{[2]}\left( \rho_.  \right) \int d j_{.,.} C^{statio}\left(  j_{.,.}  \right)
 \ e^{\displaystyle T S^{[2.5]}( \rho_. ; j_{.,.} ; a_{.,.}  )  }
 \nonumber \\
 &&  \opsimeq_{T \to +\infty} 
 C^{[2]}\left( \rho_.  \right) 
 \ e^{\displaystyle T S^{[2.25]}( \rho_.  ; a_{.,.}  )  }
\label{omegatjumpa}
\end{eqnarray}

The behavior of the even link contribution of Eq. \ref{entropyempiexplicitajlinkeven}
 with respect to the link current $j_{x,y} \in ]- a_{x,y},+a_{x,y}[$.
 can be analyzed as follows : 
 the first partial derivative with respect to the current $j_{x,y}$ 
\begin{eqnarray}
 \frac{ \partial  S_{[x,y]even}^{[2.5]} \left( \rho_x ; \rho_y ; j_{x,y} ; a_{x,y} \right)    }{ \partial j_{x,y}}
  =      - \frac{1}{2}\ln \left( \frac{ a_{x,y} + j_{x,y}  }{ a_{x,y} - j_{x,y} }  \right) 
    \label{entropyempiexplicitajlinkevenderi1}
\end{eqnarray}
is of the opposite sign of the current $j_{x,y}$, while the second partial derivative remains negative
\begin{eqnarray}
 \frac{ \partial^2  S_{[x,y]even}^{[2.5]} \left( \rho_x ; \rho_y ; j_{x,y} ; a_{x,y} \right)    }{ \partial^2 j_{x,y}}
  =  - \frac{ 2 a_{x,y}   }{ a^2_{x,y} - j^2_{x,y} } <0
    \label{entropyempiexplicitajlinkevenderi2} 
\end{eqnarray}
Since the link entropy $S_{[x,y]even}^{[2.5]} \left( \rho_x ; \rho_y ; j_{x,y} ; a_{x,y} \right)$
is maximal for vanishing current
\begin{eqnarray}
j^{opt}_{x,y}=0
\label{jopt0}
\end{eqnarray} 
the conclusion is that, for any given empirical density $\rho_.$
and any given empirical activities $a_{,.,}$,
the vanishing of the empirical currents $j^{opt}_{x,y}=0 $ on all the links $[x,y]$
allows to maximize the entropy $S^{[2.5]} \left( \rho_. ;  j_{.,.} ; a_{.,.} \right)   $ of Eq. \ref{entropyempiexplicitajintolinksevenonly} over the currents,
while the stationary constraints of Eq. \ref{cstatiojump} are trivially satisfied.
So the entropy $S^{[2.25]}( \rho_.  ; a_{.,.}  ) $ governing Eq. \ref{omegatjumpa}
is given by Eq. \ref{entropyempiexplicitajintolinksevenonly} with Eq. \ref{entropyempiexplicitajlinkeven}
 for vanishing currents $j^{opt}_{.,.}=0$
\begin{eqnarray}
S^{[2.25]}( \rho_.  ; a_{.,.}  )
&& = S^{[2.5]}( \rho_. ; j^{opt}_{.,.}=0 ; a_{.,.}  )  
=   \sum_{y  }\sum_{ x \in N^+_y  } 
           S_{[x,y]even}^{[2.5]} \left( \rho_x ; \rho_y ; j^{opt}_{x,y}=0 ; a_{x,y} \right)  
 \nonumber \\
&&           
 =   \sum_{y  }\sum_{ x \in N^+_y  }
           \left[ a_{x,y}    - \frac{a_{x,y} }{2}\ln \left( \frac{a_{x,y}^2 }{ 4 \rho_x \rho_y}  \right)      \right]
\label{entropyrhoa}
\end{eqnarray}


\subsection{ Number $\Omega^{[2.25']}_T (\rho_. ; j_{.,.}  ) $ of trajectories 
$x(.)$ with 
the empirical density $\rho_.$ and the empirical currents $j_{.,.}$   }

The number $\Omega^{[2.25']}_T (\rho_. ; j_{.,.}   ) $
of trajectories $x(0 \leq t \leq T)$ with the given empirical density $\rho_.$ and the given empirical currents $j_{.,.}$ can be obtained from the integration of Eq. \ref{omegatjumpja} over the activities $a_{.,.}$
\begin{eqnarray}
\Omega^{[2.25']}_T (\rho_. ; j_{.,.}   ) \equiv 
&& \int d a_{.,.} \Omega^{[2.5]}_T( \rho_. ; j_{.,.} ; a_{.,.} ) \opsimeq_{T \to +\infty} 
 C^{[2]}\left( \rho_.  \right) C^{statio}\left(  j_{.,.}  \right)
\int d a_{.,.}  \ e^{\displaystyle T S^{[2.5]}( \rho_. ; j_{.,.} ; a_{.,.}  )  }
\nonumber \\
&& \opsimeq_{T \to +\infty} 
 C^{[2]}\left( \rho_.  \right) C^{statio}\left(  j_{.,.}  \right)
 \ e^{\displaystyle T S^{[2.25']}( \rho_. ; j_{.,.}   )  }
\label{omegatjumpj}
\end{eqnarray}
The maximization of the entropy $S^{[2.5]}( \rho_. ; j_{.,.} ; a_{.,.}  )$ 
of Eq. \ref{entropyempiexplicitajintolinksevenonly}
over $a_{x,y}$
\begin{eqnarray}
0= \frac{ \partial S^{[2.5]} \left( \rho_. ;  j_{.,.} ; a_{.,.} \right)  }{ \partial a_{x,y} }
           =         - \frac{1 }{2}\ln \left( \frac{a_{x,y}^2 - j_{x,y}^2}{ 4 \rho_x \rho_y}  \right)    
\label{entropyempiexplicitajderia}
\end{eqnarray}
leads to the optimal values 
\begin{eqnarray}
a_{x,y}^{opt} = \sqrt{ j_{x,y}^2 + 4 \rho_x \rho_y}  
\label{activityoptimal}
\end{eqnarray}
that can be plugged into the entropy $S^{[2.5]}( \rho_. ; j_{.,.} ; a_{.,.}  )$ 
of Eq. \ref{entropyempiexplicitajintolinksevenonly}
to obtain the entropy $S^{[2.25']}( \rho_. ; j_{.,.}   )  $ at Level 2.25' governing Eq. \ref{omegatjumpj}
\begin{eqnarray}
S^{[2.25']}( \rho_. ; j_{.,.}   ) && = S^{[2.5]} \left( \rho_. ;  j_{.,.} ; a^{opt}_{.,.} \right)  
        \nonumber \\
        &&   =    
 \sum_{y  }\sum_{ x \in N^+_y  } 
\left[ \sqrt{ j_{x,y}^2 + 4 \rho_x \rho_y}  
   - \frac{ j_{x,y}}{2}\ln \left( \frac{ \sqrt{ j_{x,y}^2 + 4 \rho_x \rho_y}   + j_{x,y}  }{\sqrt{ j_{x,y}^2 + 4 \rho_x \rho_y}   - j_{x,y} }  \right)   \right]            
        \label{entropy2.25rhoj}
\end{eqnarray}


\subsection{ Number $\Omega^{[2]}_T (\rho_.  ) $ of trajectories 
$x(.)$ with the empirical density $\rho_.$ }

The number $\Omega^{[2]}_T (\rho_.  ) $ of trajectories 
$x(.)$ with the empirical density $\rho_.$ of Eq. \ref{omegalevel2}
can be computed via the integration of Eq. \ref{omegatjumpa}
over the empirical activities $a_{.,.}$
\begin{eqnarray}
\Omega^{[2]}_T (\rho_.  ) = \int da_{.,.} \Omega^{[2.25]}_T (\rho_. ; a_{.,.}   ) 
 \opsimeq_{T \to +\infty} 
 C^{[2]}\left( \rho_.  \right) 
 \int da_{.,.} e^{\displaystyle T S^{[2.25]}( \rho_.  ; a_{.,.}  )  }
  \opsimeq_{T \to +\infty} 
 C^{[2]}\left( \rho_.  \right) 
  e^{\displaystyle T S^{[2]}( \rho_.  )  }
\label{omegatjumprho}
\end{eqnarray}
The optimization of the entropy $S^{[2.25]}( \rho_.  ; a_{.,.}  ) $ of Eq. \ref{entropyrhoa}
over the activity $a_{x,y}$
\begin{eqnarray}
0= \frac{ \partial S^{[2.25]}( \rho_.  ; a_{.,.}  ) }{\partial a_{x,y}}          
 =             - \frac{1 }{2}\ln \left( \frac{a_{x,y}^2 }{ 4 \rho_x \rho_y}  \right)   
\label{entropyrhoaderi}
\end{eqnarray}
leads to the optimal values
\begin{eqnarray}
a_{x,y}^{opt} = \sqrt{  4 \rho_x \rho_y}  
\label{activityoptimalj0}
\end{eqnarray}
can be plugged into the entropy $S^{[2.25]}( \rho_.  ; a_{.,.}  )$ 
of Eq. \ref{entropyrhoa}
to obtain the explicit entropy $S^{[2]}( \rho_.  )  $ at Level 2 governing Eq. \ref{omegatjumprho}
\begin{eqnarray}
S^{[2]}( \rho_.  )  = S^{[2.25']}( \rho_.  ; a^{opt}_{.,.}  )
    =   2  \sum_{y  }\sum_{ x \in N^+_y  }\sqrt{   \rho_x \rho_y}    
    =   \sum_{y  }\sum_{ x \in N_y  }\sqrt{   \rho_x \rho_y}           
        \label{entropy2jump}
\end{eqnarray}
This entropy $S^{[2]}( \rho_.  ) $ can be also recovered as the optimal value of the entropy $S^{[2.25']}( \rho_. ; j_{.,.}  )  $ of Eq. \ref{entropy2.25rhoj}
for the optimal vanishing currents $j^{opt}_{x,y}=0$
\begin{eqnarray}
S^{[2]}( \rho_.  ) = S^{[2.25']}( \rho_. ; j^{opt}_{.,.} =0  ) 
   =    
2 \sum_{y  }\sum_{ x \in N^+_y  } 
\left[ \sqrt{   \rho_x \rho_y}  
   \right]          =   \sum_{y  }\sum_{ x \in N_y  }\sqrt{   \rho_x \rho_y}           
        \label{entropy2.25rhojvers2}
\end{eqnarray}


\subsection{ Total number $\Omega^{[0]}_T $ of trajectories 
$x(.)$  }

Let us now consider the Level 0 of Eq. \ref{omegalevel0}
via the integration of the explicit Level 2 of Eq. \ref{omegatjumprho}
over the empirical density $\rho_.$
\begin{eqnarray}
\Omega^{[0]}_T \equiv  \int d \rho_.  C^{[2]}(\rho_.) \ e^{\displaystyle   T  S^{[2]}  (\rho_.)    } \opsimeq_{T \to +\infty}    e^{\displaystyle   T  S^{[0]}   }
\label{omega0jump}
\end{eqnarray}
In order to optimize the entropy $S^{[2]}  (\rho_.)  $ at Level 2
over the empirical density $\rho_.$ in the presence of the normalization constraint $C^{[2]}(\rho_.) $
of Eq. \ref{c2jump}, let us introduce the following Lagrangian containing the Lagrange multiplier $\mu$
\begin{eqnarray}
{\cal L}_2(\rho_.) = S^{[2]}  (\rho_.)    - \mu \left( \sum_x \rho_{x} - 1 \right)
=  \sum_{y  }\sum_{ x \in N_y  }\sqrt{   \rho_x \rho_y}   - \mu \left( \sum_x \rho_{x} - 1 \right)
\label{lagrangian2jump}
\end{eqnarray}
The optimization over $\rho_x$
\begin{eqnarray}
0 = \frac{ \partial {\cal L}_2(\rho_.) }{\partial \rho_x} 
=  \sum_{ y \in N_x  } \frac{ \sqrt{    \rho_y} } { \sqrt{    \rho_x} } - \mu 
\label{lagrangian2jumpderi}
\end{eqnarray}
can be rewritten as the following eigenvalue equation for the positive eigenvector $ \sqrt{ \rho_.}$
of the symmetric neighborhood matrix satisfying $N_{x,y}=1$ if $x$ and $y$ are neighbors
\begin{eqnarray}
\mu \sqrt{   \rho_x} =  \sum_{ y \in N_x  }  \sqrt{    \rho_y}  = \sum_y N_{x,y}  \sqrt{    \rho_y} 
\label{eigenjump}
\end{eqnarray}
while $\mu$ is the corresponding highest Perron-Frobenius eigenvalue.
The entropy $S^{[0]}$ corresponds to
the optimal value of the Lagrangian obtained for the optimal density $\rho_.$
satisfying the eigenvalue Eq. \ref{eigenjump} and the normalization constraint 
\begin{eqnarray}
S^{[0]}  = {\cal L}^{opt}_2  
=  \sum_{y  } \sqrt{    \rho_y}  \left[ \sum_{ x \in N_y  }\sqrt{   \rho_x }  \right]
= \sum_{y  } \sqrt{    \rho_y}  \left[ \mu \sqrt{   \rho_y }  \right]
 = \mu  \sum_{y  }   \rho_y 
 = \mu
\label{entropy0jump}
\end{eqnarray}
The conclusion is that the entropy $S^{[0]}$ reduces to the eigenvalue $\mu$ of Eq. \ref{eigenjump},
so that it will be explicit only when the eigenvalue equation of Eq. \ref{eigenjump}
can be solved for the symmetric neighborhood matrix $N_{.,.}$ one is interested in.


\subsubsection*{ Simplest example of the finite regular lattice in dimension $d$ with periodic boundary conditions }

The simplest example is the case of a finite regular lattice in dimension $d$ with periodic boundary conditions, where each site has $(2d)$ neighbors : then the eigenvector of Eq. \ref{eigenjump}
takes the same value on each site, so that the eigenvalue reduces to $\mu=2d$ and leads to the entropy
\begin{eqnarray}
S^{[0]}   = \mu =2d
\label{entropy0jumpreg}
\end{eqnarray}
This simple value can be understood from the integration of the measure on the first line of Eq. \ref{norma},
where the sums over the $K$ positions $x_k$ simply produce the factor $(2d)^K$
\begin{eqnarray}
 \sum_{K=0}^{+\infty}  \int_0^T dt_K \int_0^{t_K} dt_{K-1} ... \int_0^{t_2} dt_{1} (2d)^K
=\sum_{K=0}^{+\infty} \frac{(2d)^K}{K!} \prod_{k=1}^K \left[ \int_0^T dt_k \right]
= \sum_{K=0}^{+\infty} \frac{(2dT)^K}{K!} =e^{2dT}
\label{normawt}
\end{eqnarray}
in agreement with the entropy of Eq. \ref{entropy0jumpreg}.


\subsection{ Number $\Omega^{[TotalActivity]}_T (A)$ of trajectories 
$x(.)$ with a given total activity $A$ }

As a simple example of additive observable, let us now consider the total activity
\begin{eqnarray}
A \equiv \sum_{y  }\sum_{ x \in N^+_y  } a_{x,y} = \sum_{y  }\sum_{ x \in N_y  } \frac{a_{x,y}}{2}
\label{totalactivity}
\end{eqnarray}
The number $\Omega^{[TotalActivity]}_T (A)$ of trajectories 
$x(.)$ with a given total activity $A$ can be obtained via the integration of
$ \Omega^{[2.25]}_T (\rho_. ; a_{.,.}   ) $ of Eq. \ref{omegatjumpa}
over the empirical density and the empirical activities with the constraint imposing the value $A$ of Eq. \ref{totalactivity}
\begin{eqnarray}
&& \Omega^{[TotalActivity]}_T (A) 
 = \int d \rho_. \int da_{.,.} 
\delta \left( A -  \sum_{y  }\sum_{ x \in N_y  } \frac{a_{x,y}}{2} \right) \Omega^{[2.25]}_T (\rho_. ; a_{.,.}   ) 
\nonumber \\
&&  \opsimeq_{T \to +\infty} 
  \int d \rho_. \int da_{.,.} \delta \left( \sum_x \rho_{x} - 1 \right) 
\delta \left( A - \sum_{y  }\sum_{ x \in N^+_y  } a_{x,y}  \right)
 \ e^{\displaystyle T S^{[2.25]}( \rho_.  ; a_{.,.}  )  }
  \opsimeq_{T \to +\infty} e^{\displaystyle T S^{[TotalActivity]}( A  )  }
\label{omegatjumpatot}
\end{eqnarray}
In order to optimize the entropy $S^{[2.25]}( \rho_.  ; a_{.,.}  ) $
over the empirical density and the empirical activities in the presence of the two constraints,
let us introduce the following Lagrangian with the two Lagrange multipliers $(\lambda,\nu)$
\begin{eqnarray}
{\cal L}_A(\rho_.  ; a_{.,.} ) 
&& = S^{[2.25]}( \rho_.  ; a_{.,.}  ) 
  - \lambda \left( \sum_x \rho_{x} - 1 \right) 
- \nu  \left(  \sum_{y  }\sum_{ x \in N^+_y  } a_{x,y} -A \right)   
\nonumber \\
&& =   \sum_{y  }\sum_{ x \in N^+_y  }
           \left[ a_{x,y}    - \frac{a_{x,y} }{2}\ln \left( \frac{a_{x,y}^2 }{ 4 \rho_x \rho_y}  \right)      \right]
             - \lambda \left( \sum_x \rho_{x} - 1 \right) 
- \nu  \left(  \sum_{y  }\sum_{ x \in N^+_y  } a_{x,y} -A \right)   
\label{lagrangianaddi}
\end{eqnarray}
The optimization over $\rho_y$
\begin{eqnarray}
0= \frac{ \partial {\cal L}_A(\rho_.  ; a_{.,.} ) }{\partial \rho_y}
 =   \frac{1}{\rho_y} \sum_{ x \in N_y  }       \frac{  a_{x,y} }{2}             - \lambda  
\label{lagrangianaddiderirho}
\end{eqnarray}
gives
\begin{eqnarray}
 \sum_{ x \in N_y  }       \frac{  a_{x,y} }{2}                 = \lambda  \rho_y
\label{lagrangianaddiderirhobis}
\end{eqnarray}
The summation over $y$ yields using the two constraints
\begin{eqnarray}
A= \sum_y     \sum_{ x \in N_y  }       \frac{  a_{x,y} }{2}           = \lambda \sum_y \rho_y = \lambda
\label{lambdasol}
\end{eqnarray}
so that the Lagrange multiplier $\lambda$ is simply given by the value $A$ of the total acitivity.
The optimization over $a_{x,y}$
\begin{eqnarray}
0= \frac{ \partial {\cal L}_A(\rho_.  ; a_{.,.} ) }{\partial a_{x,y}   }
 =       - \frac{1 }{2}\ln \left( \frac{a_{x,y}^2 }{ 4 \rho_x \rho_y}  \right)   - \nu
\label{lagrangianaddideria}
\end{eqnarray}
yields
\begin{eqnarray}
a_{x,y}= 2 e^{-\nu} \sqrt{   \rho_x \rho_y}  
\label{aoptaddi}
\end{eqnarray}
that can be plugged into Eq. \ref{lagrangianaddiderirhobis} using $\lambda=A$ of Eq. \ref{lambdasol}
\begin{eqnarray}
A  \rho_y = \sum_{ x \in N_y^+  }   2 e^{-\nu} \sqrt{   \rho_x \rho_y}    
=     e^{-\nu}  \sqrt{    \rho_y}    \sum_{ x \in N_y  }   \sqrt{   \rho_x }      
\label{lagrangianaddiderirhoter}
\end{eqnarray}
to obtain that the optimal density should satisfy
\begin{eqnarray}
\left( A     e^{\nu} \right) \sqrt{    \rho_y}  =  \sum_{ x \in N_y  }   \sqrt{   \rho_x }      
\label{lagrangianaddiderirho4}
\end{eqnarray}
where one recognizes the eigenvalue problem already discussed in Eq. \ref{eigenjump},
so that the Lagrange multiplier $\nu$ can be rewritten in terms of $A$ and in terms of the eigenvalue $\mu=S^{[0]}$ of Eq. \ref{eigenjump}
\begin{eqnarray}
\nu=  \ln \left( \frac{S^{[0]}}{A} \right)    
\label{nusol}
\end{eqnarray}
The entropy $S^{[TotalActivity]}( A  ) $ of Eq. \ref{omegatjumpatot}
corresponds to the optimal value of the entropy $S^{[2.25]}( \rho_.  ; a_{.,.}  ) $
\begin{eqnarray}
S^{[TotalActivity]}( A  )  
 = S^{[2.25]}( \rho^{opt}_.  ; a^{opt}_{.,.}  ) 
 =  A (1+\nu) = A \left[ 1+\ln \left( \frac{S^{[0]}}{A} \right)     \right] 
 \label{entropyacti}
\end{eqnarray}


\subsubsection*{ Simplest example of the finite regular lattice in dimension $d$ with periodic boundary conditions }

Let us consider again the simplest case of a finite regular lattice in dimension $d$ with periodic boundary conditions discussed around Eq. \ref{entropy0jumpreg}, so that Eq. \ref{entropyacti}
becomes
\begin{eqnarray}
S^{[TotalActivity]}( A  )   = A \left[ 1+\ln \left( \frac{2d}{A} \right)     \right] 
 \label{entropyactisimple}
\end{eqnarray}
This simple entropy function can be understood 
the number of trajectories $\frac{(2dT)^K}{K!}  $ with the total number of jumps $K=T A$ in Eq. \ref{normawt}
using the Stirling formula for the factorial $K!=(T A)!$ for large $T$
\begin{eqnarray}
 \frac{(2dT)^{TA}}{(T A)!} \opsimeq_{T \to +\infty}  \frac{(2dT)^{TA}}{\left(\frac{TA}{e} \right)^{TA}} 
 = \left( \frac{2de}{A} \right)^{TA} = e^{T A \left[ 1+\ln \left( \frac{2d}{A} \right)     \right] }
 = e^{ T S^{[TotalActivity]}( A  ) }
\label{normawtMstirling}
\end{eqnarray}
in agreement with the entropy of Eq. \ref{entropyactisimple}.


\subsection{ Number $\Omega^{[Flows]}_T (j_{.,.},a_{.,.})$ of trajectories 
$x(.)$ with the empirical currents $j_{.,.}$ and the empirical activities $a_{.,.}$}

The number $\Omega^{[Flows]}_T (j_{.,.},a_{.,.})$ of trajectories 
$x(.)$ with the empirical currents $j_{.,.}$ and the empirical activities $a_{.,.}$
can be obtained via the integration of Eq. \ref{omegatjumpja}
over the empirical density $\rho_.$
\begin{eqnarray}
\Omega^{[Flows]}_T (j_{.,.},a_{.,.}) && = \int d \rho_.  \Omega^{[2.5]}_T( \rho_. ; j_{.,.} ; a_{.,.} ) \opsimeq_{T \to +\infty} 
C^{statio}\left(  j_{.,.}  \right) \int d \rho_. 
 \delta \left( \sum_x \rho_{x} - 1 \right)
  \ e^{\displaystyle T S^{[2.5]}( \rho_. ; j_{.,.} ; a_{.,.}  )  }
  \nonumber \\
  && \opsimeq_{T \to +\infty} 
C^{statio}\left(  j_{.,.}  \right) 
  \ e^{\displaystyle T S^{[Flows]}(  j_{.,.} ; a_{.,.}  )  }
\label{omegatjumpflows}
\end{eqnarray}
where the entropy of Eq. \ref{entropyempiexplicitajintolinksevenonly}
can be rewritten 
as
\begin{eqnarray}
S^{[2.5]} \left( \rho_. ;  j_{.,.} ; a_{.,.} \right)  
&&           =   \sum_{y  }\sum_{ x \in N^+_y  } 
\left[ a_{x,y}  
   - \frac{a_{x,y} }{2}\ln \left( \frac{a_{x,y}^2 - j_{x,y}^2}{ 4 }  \right)    
   - \frac{ j_{x,y}}{2}\ln \left( \frac{ a_{x,y} + j_{x,y}  }{ a_{x,y} - j_{x,y} }  \right)   \right]       
  +  \sum_{y  }\sum_{ x \in N^+_y  }  \frac{a_{x,y} }{2} \left[ \ln ( \rho_x ) + \ln ( \rho_y ) \right]   
  \nonumber \\
&& = \sum_{y  }\sum_{ x \in N^+_y  } 
\left[ a_{x,y}  
   - \frac{a_{x,y} }{2}\ln \left( \frac{a_{x,y}^2 - j_{x,y}^2}{ 4 }  \right)    
   - \frac{ j_{x,y}}{2}\ln \left( \frac{ a_{x,y} + j_{x,y}  }{ a_{x,y} - j_{x,y} }  \right)   \right]       
  +  \sum_{x  } A_x \ln (\rho_x)        
                    \label{entropy2.5forrho}
\end{eqnarray}
where we have introduced the notation
\begin{eqnarray}
A_x \equiv \sum_{ y \in N_x  } \frac{a_{x,y}}{2}
\label{Ax}
\end{eqnarray}
for the activity related to the point $x$, while the total activity of Eq. \ref{totalactivity}
studied in the previous subsection corresponds to
\begin{eqnarray}
A  =\sum_x \sum_{ y \in N_x  } \frac{a_{x,y}}{2} =\sum_{x  } A_x
\label{totalAsumx}
\end{eqnarray}
In order to optimize the entropy $S^{[2.5]} \left( \rho_. ;  j_{.,.} ; a_{.,.} \right)  $ over the empirical density $\rho_.$ satisfying the normalization constraint,
one introduce the following Lagrangian containing the Lagrange multiplier $\eta$
\begin{eqnarray}
&& {\cal L}_{[j_{.,.},a_{.,.}]}(\rho_.  ) 
 = S^{[2.5]} \left( \rho_. ;  j_{.,.} ; a_{.,.} \right) 
  - \eta \left( \sum_x \rho_{x} - 1 \right) 
\nonumber \\
&& =   \sum_{y  }\sum_{ x \in N^+_y  } 
\left[ a_{x,y}  
   - \frac{a_{x,y} }{2}\ln \left( \frac{a_{x,y}^2 - j_{x,y}^2}{ 4 }  \right)    
   - \frac{ j_{x,y}}{2}\ln \left( \frac{ a_{x,y} + j_{x,y}  }{ a_{x,y} - j_{x,y} }  \right)   \right]       
  +  \sum_{x  } A_x \ln (\rho_x)       
             - \eta \left( \sum_x \rho_{x} - 1 \right) 
\label{lagrangianrho}
\end{eqnarray}
The optimization over $\rho_x$
\begin{eqnarray}
0= \frac{ \partial {\cal L}_{[j_{.,.},a_{.,.}]}(\rho_.  ) }{\partial \rho_x}
 =   \frac{A_x}{\rho_x}         - \eta  
\label{lagrangianrhoderi}
\end{eqnarray}
leads to the optimal value
\begin{eqnarray}
 \rho_x       = \frac{A_x}{\eta}
\label{ehoxopteta}
\end{eqnarray}
The summation over $x$ yields using the normalization 
\begin{eqnarray}
1=\sum_x \rho_x = \frac{1}{\eta}\sum_x A_x = \frac{A}{\eta} 
\label{etasolsol}
\end{eqnarray}
so that the Lagrange multiplier $\eta$ is simply given by the total acitivity $A$.

Plugging the optimal solution
\begin{eqnarray}
 \rho_x       = \frac{A_x}{A}  
\label{ehoxoptetaA}
\end{eqnarray}
into $ S^{[2.5]} \left( \rho_. ;  j_{.,.} ; a_{.,.} \right) $ of Eq. \ref{entropy2.5forrho}
yields the entropy $S^{[Flows]}(  j_{.,.} ; a_{.,.}  )  $ 
governing Eq. \ref{omegatjumpja}
\begin{eqnarray}
&& S^{[Flows]}(  j_{.,.} ; a_{.,.}  )    = S^{[2.5]} \left( \rho^{opt}_. ;  j_{.,.} ; a_{.,.} \right)  
\nonumber \\
&& = \sum_{y  }\sum_{ x \in N^+_y  } 
\left[ a_{x,y}  
   - \frac{a_{x,y} }{2}\ln \left( \frac{a_{x,y}^2 - j_{x,y}^2}{ 4 }  \right)    
   - \frac{ j_{x,y}}{2}\ln \left( \frac{ a_{x,y} + j_{x,y}  }{ a_{x,y} - j_{x,y} }  \right)   \right]       
  +  \sum_{y  } A_y \ln \left(  \frac{A_y}{A}\right)    
  \nonumber \\
&& = \sum_{y  }\sum_{ x \in N^+_y  } 
\left[ a_{x,y}  
   - \frac{a_{x,y} }{2}\ln \left( \frac{a_{x,y}^2 - j_{x,y}^2}{ 4 }  \right)    
   - \frac{ j_{x,y}}{2}\ln \left( \frac{ a_{x,y} + j_{x,y}  }{ a_{x,y} - j_{x,y} }  \right)   \right]       
  +  \sum_{y  } 
  \sum_{ x \in N_y  } \frac{a_{x,y}}{2}  
   \ln \left(  \frac{ \displaystyle \sum_{ z \in N_y  } \frac{a_{z,y}}{2}}
   {  \displaystyle \sum_{z'} \sum_{ z'' \in N_{z'}  } \frac{a_{z',z''}}{2} }\right)        \ \    
                    \label{entropyflowsja}
\end{eqnarray}
where we have replaced $A_y$ and $A$ by their expressions of Eqs \ref{Ax}
and \ref{totalAsumx}
in terms of the link activities $a_{.,.}$.


\subsection{ Number $\Omega^{[Activities]}_T (a_{.,.})$ of trajectories 
$x(.)$ with the empirical activities $a_{.,.}$}

The number $\Omega^{[Activities]}_T (a_{.,.})$ of trajectories 
$x(.)$ with the empirical activities $a_{.,.}$
can be obtained via the integration of Eq. \ref{omegatjumpflows}
over the empirical currents $j_{.,.}$
\begin{eqnarray}
\Omega^{[Activities]}_T (a_{.,.}) && = \int dj_{.,.} \Omega^{[Flows]}_T (j_{.,.},a_{.,.}) 
 \opsimeq_{T \to +\infty} \int dj_{.,.} 
C^{statio}\left(  j_{.,.}  \right) 
  \ e^{\displaystyle T S^{[Flows]}(  j_{.,.} ; a_{.,.}  )  }
  \nonumber \\
  && \opsimeq_{T \to +\infty} 
  \ e^{\displaystyle T S^{[Activities]}( a_{.,.}  )  }
\label{omegatjumpactivities}
\end{eqnarray}
As in Eq. \ref{jopt0}, one obtains that the optimal currents vanish
\begin{eqnarray}
j^{opt}_{x,y}=0
\label{jopt0bia}
\end{eqnarray} 
so that the entropy $S^{[Activities]}( a_{.,.}  ) $ is directly obtained from the entropy of Eq. \ref{entropyflowsja}
\begin{eqnarray}
S^{[Activities]}( a_{.,.}  ) && =S^{[Flows]}(  j^{opt}_{.,.}=0 ; a_{.,.}  )   
\nonumber \\
&& = \sum_{y  }\sum_{ x \in N_y  } 
\left[ \frac{ a_{x,y}  }{2}
   - \frac{ a_{x,y}  }{2} \ln \left( \frac{a_{x,y} }{ 2 }  \right)    
   \right]       
  +  \sum_{y  } 
  \sum_{ x \in N_y  } \frac{a_{x,y}}{2}  
   \ln \left(  \frac{ \displaystyle \sum_{ z \in N_y  } \frac{a_{z,y}}{2}}
   {  \displaystyle \sum_{z'} \sum_{ z'' \in N_{z'}  } \frac{a_{z',z''}}{2} }\right)          
                    \label{entropyactivities}
\end{eqnarray}



\section{ Conclusions  }

\label{sec_conclusion}

In this paper, we have revisited the statistical physics of Markov trajectories $x(0 \leq t \leq T)$
via the notion of the Canonical Ensemble at Level 2.5 associated to the Markov generator $M$
and the notion of the Microcanonical Ensemble at Level 2.5 associated to fixed values of all the relevant empirical observables $E_n$.

We have first explained why the Ensemble of trajectories $x(0 \leq t \leq T)$ produced by the Markov generator $M$ can be considered as 'Canonical' :

(C1) The probability of the trajectory $x(0 \leq t \leq T)$ can be rewritten as the exponential of a linear combination of its relevant empirical observables $E_n$, where the coefficients involving the Markov generator are their fixed conjugate parameters.

(C2) The large deviations properties of these relevant empirical observables $E_n$ for large $T$ are governed by the explicit rate function $I^{[2.5]}_M  (E_.) $ at Level 2.5, while in the thermodynamic limit $T=+\infty$, they concentrate on their typical values $E_n^{typ[M]}$ determined by the Markov generator $M$.

We have then analyzed the properties of the 'Microcanonical Ensemble' at Level 2.5 for stochastic trajectories $x(0 \leq t \leq T)$, where all the relevant empirical variables $E_n$ are fixed to some values $E^*_n$ and cannot fluctuate anymore for finite $T$ :

(MC1) When the long trajectory $x(0 \leq t \leq T) $ belongs the Microcanonical Ensemble at Level 2.5 with the fixed empirical observables $E_n^*$, the statistics of its subtrajectory $x(0 \leq t \leq \tau)  $ for $1 \ll \tau \ll T $ is governed by the Canonical Ensemble at Level 2.5 associated to the Markov generator $M^*$ that would make the empirical observables $E_n^*$ typical.

(MC2) In the Microcanonical Ensemble at Level 2.5, the central role is played by the number $\Omega^{[2.5]}_T(E^*_.) $ of stochastic trajectories of duration $T$ with the given empirical observables $E^*_n$, and by the corresponding explicit Boltzmann entropy $S^{[2.5]}( E^*_. )  = [\ln \Omega^{[2.5]}_T(E^*_.)]/T  $ at Level 2.5.

We have described in detail how this general framework can be applied to continuous-time Markov Jump processes and to discrete-time Markov chains, with the simple examples of directed trajectories on a ring.
Finally for the special case of undirected Markov Jump processes (where the jumps between two configurations are either both possible or both impossible), we have shown 
how the entropy $S^{[2.5]}( \rho_., j_{.,.}, a_{.,.}  ) $ at Level 2.5 as a function of the empirical density $\rho_.$, the empirical currents $j_{.,.}$ and the empirical activities $a_{.,.}$ can be contracted to obtain the explicit entropies of many other lower levels.


 \appendix

\section{ Application to discrete-time Markov chains  }

\label{app_chain}

In this Appendix, we describe how the general framework of section \ref{sec_general}
can be applied to discrete-time Markov chains in a discrete configuration space.

\subsection{ Canonical Ensemble of trajectories $x(0 \leq t \leq T)$ associated to a Markov Chain generator $W$   }

 \subsubsection{Discrete-time Markov chain converging towards some normalizable steady state} 
 
Let us now consider the discrete-time Markov chain dynamics 
for the probability $P_y(t)  $ to be in configuration $y$ at time $t$
\begin{eqnarray}
P_x(t+1) =  \sum_y W_{x,y}  P_y(t)
\label{markovchain}
\end{eqnarray}
where the Markov Matrix elements are positive $W_{x,y} \geq 0 $ and satisfy the normalization
\begin{eqnarray}
  \sum_x W_{x,y} && =1
\label{markovnorma}
\end{eqnarray}
The steady-state solution $P^*_x $ of Eq. \ref{markovchain}
\begin{eqnarray}
P^*_x =  \sum_y W_{x,y} P^*_y
\label{markovchainst}
\end{eqnarray}
is assumed to be normalizable
\begin{eqnarray}
1 =    \sum_{y }  P_y^* 
\label{normastc}
\end{eqnarray}
 

\subsubsection{Identification of the relevant empirical observables that determine the trajectories probabilities } 

 The probability of the whole trajectory $x(0 \leq t \leq T)$ starting at the fixed configuration $x_0$ at time $t=0$
 reads
\begin{eqnarray}
{\cal P}[x(0 \leq t \leq T)]  =   \delta_{x(0),x_0 } \left[ \prod_{t=1}^T W_{x(t) ,x(t-1)} \right] 
\label{pwtrajchain}
\end{eqnarray}

The corresponding information per unit time ${\cal I}_W\left[ x(0 \leq t \leq T) \right] $
of Eq. \ref{informationdef}
reads 
\begin{eqnarray}
{\cal I}_W\left[ x(.) \right]  \equiv - \frac{{\cal P}_{[W_{.,.}]}\left[ x(.) \right] }{T}
=  - \frac{1}{T}   \sum_{t=1}^T \ln \left(W_{x(t) ,x(t-1)} \right)
\label{informationchain}
\end{eqnarray}
so that it depends only the empirical time-averaged 2-point density
 \begin{eqnarray}
 \rho_{x,y}  \equiv \frac{1}{T} \sum_{t=1}^T \delta_{x(t),x}  \delta_{x(t-1),y} 
\label{rho2pt}
\end{eqnarray}
 The empirical 2-point density allows to reconstruct the empirical 1-point density
via the summation over the first or the second index 
 (up to a boundary term of order $1/T$ that is negligible for large time $T \to +\infty$)
\begin{eqnarray}
 \rho_{x}  \equiv \frac{1}{T} \sum_{t=1}^T \delta_{x(t),x}  = \sum_y \rho_{x,y} =
 \sum_y \rho_{y,x}
\label{rho1pt}
\end{eqnarray}
with the normalization
\begin{eqnarray}
\sum_x \rho_{x}  = 1
\label{rho1ptnorma}
\end{eqnarray}
 
 The typical value of the empirical 1-point density is the steady state of Eq. \ref{markovchainst}
\begin{eqnarray}
 \rho^{typ[W]}_{x} = P^*_x 
\label{rho1pttyp}
\end{eqnarray}
while the typical value of the empirical 2-point density is given by the steady-state flows
of Eq. \ref{markovchainst}
\begin{eqnarray}
 \rho^{typ[W]}_{x,y} = W_{x,y}  P^*_y  = W_{x,y} \rho^{typ[W]}_y = W_{x,y} \sum_z \rho^{typ[W]}_{z,y}
\label{rho2pttyp}
\end{eqnarray}

 With respect to the general formalism of section \ref{sec_general},
this means that 

(i) the relevant empirical observables $E_n$ that determine the trajectories probabilities
are given by the empirical 2-point 
density $\rho_{x,y}$ 
\begin{eqnarray}
 E_. = \left( \rho_{.,.} \right)
\label{Echain}
\end{eqnarray}

(ii)  their constitutive constraints $C^{[2.5]}(E_.)$ are given by Eqs \ref{rho1pt} and \ref{rho1ptnorma}
\begin{eqnarray}
 C^{[2.5]}\left( \rho_{.,.} \right)
  = \delta \left( \sum_{x,y} \rho_{x,y} - 1 \right) 
 \prod_x \left[ \delta \left(  \sum_y \rho_{x,y}  - \sum_y \rho_{y,x}  \right)\right]
\label{constraints2.5chain}
\end{eqnarray}

(iii) the information $I_W(E_.)$ obtained from Eq. \ref{informationchain}
corresponds to the following linear combination of the empirical observables $E_.=\left( \rho_{.,.} \right)$  
\begin{eqnarray}
I_W \left( \rho_{.,.} \right) = - \sum_{x,y}  \rho_{x,y} \ln \left(W_{x ,y} \right) 
\equiv \sum_{x,y}  \rho_{x,y} i_{x,y} (W)
\label{actionchain}
\end{eqnarray}
where the coefficients representing their intensive congugate parameters
\begin{eqnarray}
i_{ x,y } (W) \equiv -  \ln \left(W_{x ,y} \right)
\label{lambdachain}
\end{eqnarray}
are very simple in terms of the Markov matrix elements $W_{x ,y} $.


\subsubsection{ Boltzmann intensive entropy $S^{[2.5]}\left( \rho_{.,.} \right)$ as a function of the empirical 2-point density $ \rho_{.,.} $ }

Eq. \ref{rho2pttyp} yields that the modified Markov matrix $W^E$ that would make
the empirical observables $E_.=\left( \rho_{.,.} \right)$ typical 
is given by 
\begin{eqnarray}
W^E_{x ,y} \equiv \frac{\rho_{x,y}}{ \rho_{y}} = \frac{\rho_{x,y}}{ \sum_z \rho_{z,y}}
\label{wchaininfer}
\end{eqnarray}
As a consequence,
Eq. \ref{compensation} yields that the 
entropy $S^{[2.5]}\left( \rho_{.,.} \right)$ as a function of the empirical observables $E_.=\left( \rho_{.,.} \right)$ reads using Eqs \ref{actionchain} and \ref{wchaininfer}
\begin{eqnarray}
S^{[2.5]} \left( \rho_{.,.} \right)    =  I_{W^E}  \left( \rho_{.,.} \right)   
  =    - \sum_{x,y}  \rho_{x,y} \ln \left( \frac{\rho_{x,y}}{ \rho_{y}}  \right)
  = - \sum_{x,y}  \rho_{x,y} \ln \left( \frac{\rho_{x,y}}{ \sum_z \rho_{z,y}}  \right)
\label{entropyempichain}
\end{eqnarray}
As stressed after Eq. \ref{entropynonlinear}, it is non-linear with respect to the empirical observables 
$E_.=\left( \rho_{.,.} \right)$.

\subsubsection{ Rate function $I^{[2.5]}_W( \rho_{.,.} ) $ at level 2.5 for the empirical observables $E_.=\left( \rho_{.,.} \right)$ }

For large $T$, the probability to see the empirical 2-point density $\rho_{.,.}  $
follows the large deviation form at Level 2.5   \cite{fortelle_thesis,fortelle_chain,review_touchette,c_largedevdisorder,c_reset,c_inference}
\begin{eqnarray}
P^{[2.5]}_W ( \rho_{.,.}  )
  \opsimeq_{T \to +\infty}  C^{[2.5]} ( \rho_{.,.} )e^{ - T I^{[2.5]}_W( \rho_{.,.} )  } 
\label{level2.5chain}
\end{eqnarray}
where the constitutive constraints $C ( \rho_{.,.} )$ have been written in Eq. \ref{constraints2.5chain},
while the rate function 
 is given by the difference between the information 
of Eq. \ref{actionchain}
and the entropy from Eq. \ref{entropyempichain}
\begin{eqnarray}
  I^{[2.5]}_W( \rho_{.,.} ; \rho_. ) 
  =    I_W \left( \rho_{.,.} \right) - S^{[2.5]} \left( \rho_{.,.} \right)
= \sum_x \sum_y \rho_{x,y} \ln \left( \frac{\rho_{x,y}}{ W_{x,y}  \sum_z \rho_{z,y}}  \right) 
\label{rate2.5chain}
\end{eqnarray}


\subsubsection{ Kolmogorov-Sinai entropy $h_W^{KS}$ }

The Kolmogorov-Sinai entropy $h^{KS}_w$ of Eqs \ref{hksempityp} and \ref{hksempitypSalso}
reads using the typical values of Eqs \ref{rho2pttyp}
\begin{eqnarray}
h^{KS}_W  =   - \sum_{x,y}  \rho^{typ[W]}_{x,y} \ln \left(W_{x ,y} \right)
=  - \sum_{x,y}  W_{x,y} P^*_y \ln \left(W_{x ,y} \right)
\label{hksempitypchain}
\end{eqnarray}
so that it can be thus evaluated whenever the steady state $P^*(y)$ associated to the Markov matrix $W$ is known
(see \cite{c_ruelle} for simple examples in relation with the Ruelle thermodynamic formalism).


\subsection{ Microcanonical Ensemble at Level 2.5 with fixed empirical 2-point density $\rho^*_{.,.}$ }

\subsubsection{ Microcanonical Ensemble at Level 2.5 where the empirical 2-point density $\rho^*_{.,.}$ cannot fluctuate for finite $T$}
 
In the Microcanonical Ensemble of Eq. \ref{microcanodeltaempirical}
 \begin{eqnarray}
P^{Micro[2.5]}_{\left( \rho^*_{.,.} \right)} \left( \rho_{.,.} \right)    
 =  \prod_{x,y}  \delta \left( \rho_{x,y} - \rho_{x,y}^* \right) 
\label{microempichain}
\end{eqnarray}
the empirical 2-point density $\rho_{.,.} $ is fixed to 
$ \rho^{*}_{.,.} $ satisfying the constitutive constraints of Eq. \ref{constraints2.5chain}
\begin{eqnarray}
 C^{[2.5]}\left( \rho^*_{.,.} \right)
 =  \delta \left( \sum_{x,y} \rho^*_{x,y} - 1 \right) 
 \prod_x \left[ \delta \left(  \sum_y \rho^*_{x,y}  - \sum_y \rho^*_{y,x}  \right)\right]
\label{constraints2.5etoilechain}
\end{eqnarray}
and cannot fluctuate for finite $T$, in contrast to Eq. \ref{level2.5chain} 
concerning the Level 2.5 in the Canonical Ensemble associated to the Markov matrix $W$.

 
\subsubsection{ Probabilities of trajectories $x(0 \leq t \leq T) $ in the Microcanonical Ensemble  }

In the Microcanonical Ensemble of Eq. \ref{ptrajmicro}, only the trajectories $[x(0 \leq t \leq T)]  $
corresponding to the empirical 2-point density $ \rho^{*}_{.,.} $ have a non-zero weight,
and all these allowed trajectories have the same weight given by the inverse of their number $\Omega^{[2.5]}_T\left( \rho^*_{.,.} \right) $ of Eq. \ref{omegaempi}
\begin{eqnarray}
{\cal P}^{Micro[2.5]}_{\left( \rho^*{.,.} \right)}[x(.)]   
 =  \frac{ 1 }{\Omega^{[2.5]}_T\left( \rho^*_{.,.} \right) }\prod_{x,y} \delta \left( \rho^*_{x,y} - \rho_{x,y} \left[ x(.) \right] \right)
\label{ptrajmicrochain}
\end{eqnarray}
For large $T$, Eq. \ref{ptrajmicro} reads
\begin{eqnarray}
{\cal P}^{Micro[2.5]}_{\left( \rho^*_{.,.} \right)}[x(.)]   
&& \opsimeq_{T \to +\infty} 
 e^{\displaystyle - T  S^{[2.5]} \left( \rho^*_{.,.} \right)  }\prod_{x,y} \delta \left( \rho^*_{x,y} - \rho_{x,y} \left[ x(.) \right] \right)
\label{ptrajmicrotchain}
\end{eqnarray}
with the entropy $S^{[2.5]}\left( \rho^*_{.,.} \right) $ of Eq. \ref{entropyempichain}
\begin{eqnarray}
S^{[2.5]} \left( \rho^*_{.,.} \right)  
  =   - \sum_{x,y}  \rho^*_{x,y} \ln \left( \frac{\rho^*_{x,y}}{ \sum_z \rho^*_{z,y}}  \right)
\label{entropyempistarchain}
\end{eqnarray}


\subsubsection{ Statistics of the subtrajectories on $[0,\tau]$ of the Microcanonical Ensemble trajectories on $[0,T]$ for $1 \ll \tau \ll T$ }

As explained in details in subsection \ref{subsec_subtraj},
the fact that the Canonical Ensemble emerges to describe the statistics
of the subtrajectories is based on the property
of Eq. \ref{derientropyempirical} 
concerning  
the derivatives of the entropy $S^{[2.5]}\left(E_. \right) $ with respect to the empirical observables $E_n$.

For Markov chains, the derivative of the entropy $S^{[2.5]} \left( \rho_{.,.} \right) $ of Eq. \ref{entropyempichain}
with respect to the empirical 2-point density $\rho_{x,y}  $
indeed involves the coefficients $i_{ x,y } (W^E) $ of Eq. \ref{lambdachain}
associated to the modified Markov matrix $W^E_{x ,y} $ of Eq. \ref{wchaininfer}
\begin{eqnarray}
\frac{ \partial S^{[2.5]} \left( \rho_{.,.} \right)  }{\partial \rho_{x,y}}   
  = -  \ln \left( \frac{\rho_{x,y}}{ \sum_z \rho_{z,y}}  \right) -1+1
=  -  \ln \left( \frac{\rho_{x,y}}{ \sum_z \rho_{z,y}}  \right) = - \ln \left( W^E_{x ,y}\right)
= i_{ x,y } (W^E)
\label{entropyempichainderi}
\end{eqnarray}
in agreement with the general property of Eq. \ref{derientropyempirical}.


\subsection{Simple example of discrete-time directed trajectories on a ring : entropies at various levels  }

The simplest example of directed trajectories is based on
a ring of $L$ sites with periodic boundary conditions $x+L \equiv x$ :
when the particle is on site $y$ at time $t$,
it can either jump to the right neighbor $(y+1)$ 
or it remains on site $y$, so that the empirical 2-point density $\rho_{x,y}$ is non-vanishing only for $x=y$ and $x=y+1$
\begin{eqnarray}
\rho_{x,y} = \delta_{x,y} \rho_{y,y} + \delta_{x,y+1} \rho_{y+1,y}
\label{rho2trap}
\end{eqnarray}

\subsubsection{ Parametrization of the 2-point empirical density $\rho_{.,.}$ in terms of the 1-point density $\rho_.$ and of the global current $J$ } 

The empirical 1-point density $\rho_.$
can be computed from the empirical 2-point density $\rho_{x,y}$ of Eq. \ref{rho2trap}
 via the two possible sums of Eq. \ref{rho1pt}
\begin{eqnarray}
 \rho_{x}  &&  = \sum_y \rho_{x,y} =  \rho_{x,x} +  \rho_{x,x-1}
 \nonumber \\
 \rho_{x}  && =  \sum_y \rho_{y,x} =   \rho_{x,x} +  \rho_{x+1,x}
\label{rho1trap}
\end{eqnarray}
The compatibility between the two equations yields that the $L$ elements $ \rho_{x+1,x} $
take the same value $J$ along the ring $x=1,2,..,L$, that represents the current $J$
flowing through each link of the ring
\begin{eqnarray}
  \rho_{x+1,x} = J
\label{Jtrap}
\end{eqnarray}
The remaining diagonal elements $\rho_{x,x}$ can be then computed 
 from the 1-point density $\rho_x$ and the current $J$ via Eq. \ref{rho1trap}
 \begin{eqnarray}
  \rho_{x,x} = \rho_{x}  - J
\label{rho2trapdiag}
\end{eqnarray}
In summary, the 2-point empirical density of Eq. \ref{rho2trap} is now parametrized by
\begin{eqnarray}
\rho_{x,y} = \delta_{x,y} \left( \rho_{x}  - J \right) + \delta_{x,y+1} J
\label{rho2trappara}
\end{eqnarray}
where the current $J$ shoud be positive $J \geq 0$ and smaller $J \leq \rho_x$
than the empirical density $\rho_x$ for any $x=1,..,L$.
In summary, the remaining constitutive constraints read
\begin{eqnarray}
C^{[2.5]} \left( \rho_{.} ,J \right)  
  =   \delta \left( \sum_{x=1}^L \rho_x -1\right) \theta(J) \left[ \prod_{x=1}^L \theta( \rho_x - J)\right]
  \label{c2.5trap}
\end{eqnarray}

\subsubsection{ Number of Trajectories $\Omega^{[2.5]}_T\left( \rho_{.} ,J\right) $ with the empirical density $\rho_.$ and the current $J$ } 

Via the parametrization of Eq. \ref{rho2trappara}, 
the entropy of Eq. \ref{entropyempichain} 
reduces to
\begin{eqnarray}
S^{[2.5]} \left( \rho_{.} ,J \right)  
 &&  =    \sum_{x=1}^L \left[ -  \left( \rho_{x}  - J \right) \ln \left( \frac{ \rho_{x}  - J}{ \rho_x}  \right)
   -  J \ln \left( \frac{J}{ \rho_x}  \right) \right]
 \nonumber \\
 &&  =  \sum_{x=1}^L \left[ -   \rho_{x}   \ln \left( \frac{ \rho_{x}  - J}{ \rho_x}  \right)
   +  J \ln \left( \frac{\rho_{x}  - J}{ J}  \right) \right]
\label{entropy2.5trap}
\end{eqnarray}
and
the number of trajectories $\Omega^{[2.5]}_T\left( \rho_{.} ,J\right) $ 
 reads
\begin{eqnarray}
\Omega^{[2.5]}_T\left( \rho_{.} ,J\right)
 \opsimeq_{T \to +\infty} C^{[2.5]} \left( \rho_{.} ,J \right)  
 e^{\displaystyle  T  S^{[2.5]} \left( \rho_{.} ,J \right)  }
 \label{trapomega2.5}
\end{eqnarray}

\subsubsection{ Number of Trajectories $\Omega^{[2]}_T\left( \rho_{.} \right) $ with the empirical density $\rho_.$  } 

The number of trajectories $\Omega^{[2]}_T\left( \rho_{.} \right) $ 
with the empirical density $\rho_.$ can be obtained via the integration 
of Eq. \ref{trapomega2.5} over the current $J$
\begin{eqnarray}
\Omega^{[2]}_T\left( \rho_{.} \right) && = \int dJ \Omega^{[2.5]}_T\left( \rho_{.} ,J\right)
 \opsimeq_{T \to +\infty} 
   \delta \left( \sum_{x=1}^L \rho_x -1\right) 
   \int_0^{+\infty} dJ  \left[ \prod_{x=1}^L \theta( \rho_x - J)\right]
 e^{\displaystyle  T  S^{[2.5]} \left( \rho_{.} ,J \right)  }
\nonumber \\
&&   \opsimeq_{T \to +\infty} 
   \delta \left( \sum_{x=1}^L \rho_x -1\right) 
   e^{\displaystyle  T  S^{[2]} \left( \rho_{.}  \right)  }
 \label{trapomega2}
\end{eqnarray}
The optimization of the entropy $S^{[2.5]} \left( \rho_{.} ,J \right) $ at Level 2.5 over the current $J$
\begin{eqnarray}
0 = \frac{ \partial S^{[2.5]} \left( \rho_{.} ,J \right)  }{ \partial J}
  =    \sum_{x=1}^L  \ln \left( \frac{ \rho_{x}  - J}{ J }  \right) 
  = \ln \left[ \prod_{x=1}^L \left( \frac{ \rho_{x}  - J}{ J } \right) \right]
\label{entropy2.5trapderiJ}
\end{eqnarray}
yields that the optimal current $J^{opt}=J^{opt}[\rho_.]$ as a function of the given empirical density $\rho_.$
is given by the solution of
\begin{eqnarray}
1=  \prod_{x=1}^L \left( \frac{ \rho_{x}  - J^{opt}}{ J^{opt}  } \right)
\label{trapjopt}
\end{eqnarray}
It should be plugged into the entropy at Level 2.5
to obtain the entropy at Level 2
\begin{eqnarray}
S^{[2]} \left( \rho_{.}  \right) = S^{[2.5]} \left( \rho_{.} ,J^{opt}[\rho_.] \right)  
  =  -  \sum_{x=1}^L    \rho_{x}   \ln \left( \frac{ \rho_{x}  - J^{opt}[\rho_.] }{\rho_x} \right)
\label{entropy2trap}
\end{eqnarray}

\subsubsection{ Number of Trajectories $\Omega^{[Current]}_T\left( J\right) $ 
with the empirical current $J$  } 

The number of trajectories $\Omega^{[Current]}_T\left( J\right) $ 
with the empirical current $J$  can be obtained via the integration 
of Eq. \ref{trapomega2.5} over the empirical density $\rho_.$
\begin{eqnarray}
\Omega^{[Current]}_T\left( J\right)  && = \int d\rho_. \Omega^{[2.5]}_T\left( \rho_{.} ,J\right)
 \opsimeq_{T \to +\infty} \theta(J) \int d\rho_.
   \delta \left( \sum_{x=1}^L \rho_x -1\right) 
  \left[ \prod_{x=1}^L \theta( \rho_x - J)\right]
 e^{\displaystyle  T  S^{[2.5]} \left( \rho_{.} ,J \right)  }
 \label{trapomegaJ}
\end{eqnarray}
To optimize the entropy $S^{[2.5]} \left( \rho_{.} ,J \right) $ over the empirical density $\rho_.$
satisfying the normalization constraint, one introduces the following Lagrangian containing the Lagrange multiplier $\mu$
\begin{eqnarray}
{\cal L}_J(\rho_.) && \equiv S^{[2.5]} \left( \rho_{.} ,J \right)  -\mu \left( \sum_{x=1}^L \rho_x -1\right) 
\nonumber \\
 &&  =  \sum_{x=1}^L \left[ -   \rho_{x}   \ln \left( \frac{ \rho_{x}  - J}{ \rho_x}  \right)
   +  J \ln \left( \frac{\rho_{x}  - J}{ J}  \right) \right] -\mu \left( \sum_{x=1}^L \rho_x -1\right)
\label{traplagrangian}
\end{eqnarray}
The optimization over $\rho_x$
\begin{eqnarray}
0=\frac{ \partial {\cal L}_J(\rho_.)}{\partial \rho_x}     =  
 -     \ln \left( \frac{ \rho_{x}  - J}{ \rho_x}  \right)
   -\mu 
\label{traplagrangianderi}
\end{eqnarray}
yields that the optimal density $\rho_x^{opt}$ is a function of $J$ and $\mu$,
so that it does not depend on the position $x$, and its value is thus fixed by the normalization constraint
\begin{eqnarray}
 \rho_x^{opt} = \frac{1}{L}
\label{traprhounifchain}
\end{eqnarray}
that can be plugged into the entropy at Level 2.5
to obtain the entropy $S^{[Current]}\left( J\right) $ for the current alone
\begin{eqnarray}
S^{[Current]}\left( J\right)=S^{[2.5]} \left( \rho^{opt}_{.} ,J \right)  
 = - (1-LJ) \ln (1-LJ) - LJ \ln (LJ)
\label{entropyJtrap}
\end{eqnarray}
that governs Eq. \ref{trapomegaJ}
\begin{eqnarray}
\Omega^{[Current]}_T\left( J\right)  && = \int d\rho_. \Omega^{[2.5]}_T\left( \rho_{.} ,J\right)
 \opsimeq_{T \to +\infty} \theta(J)  \theta\left( \frac{1}{L} - J\right)
 e^{\displaystyle  T  S^{[Current]}\left( J\right)  }
 \label{trapomegaJfinal}
\end{eqnarray}

\subsubsection{ Total number of Trajectories $\Omega^{[0]}_T $ at Level 0  } 

The total number $\Omega^{[0]}_T $ of trajectories can be obtained from the integration
of Eq. \ref{trapomegaJfinal} over the current $J$
\begin{eqnarray}
\Omega^{[0]}_T = \int dJ \Omega^{[Current]}_T\left( J\right)  
 \opsimeq_{T \to +\infty} \int_0^{\frac{1}{L}} dJ
 e^{\displaystyle  T  S^{[Current]}\left( J\right)  } \opsimeq_{T \to +\infty}e^{\displaystyle  T  S^{[0]}  }
 \label{trapomega0chain}
\end{eqnarray}
The optimization of the entropy $S^{[Current]}\left( J\right) $ of Eq. \ref{entropyJtrap}
over the current $J$
\begin{eqnarray}
0= \frac{ \partial S^{[Current]}\left( J\right) }{\partial J} 
 =L \ln \left( \frac{1-LJ }{LJ} \right)
\label{entropyJtrapderi}
\end{eqnarray}
yields the optimal value
\begin{eqnarray}
J^{opt} = \frac{1}{2L}
\label{Joptzero}
\end{eqnarray}
that can be plugged into the entropy $S^{[Current]}\left( J\right) $ of Eq. \ref{entropyJtrap}
to obtain the entropy at Level 0 
\begin{eqnarray}
S^{[0]} = S^{[Current]}\left( J^{opt}\right)  
 = \ln 2
\label{entropytrap0}
\end{eqnarray}
i.e. one recovers that the number of trajectories of Eq. \ref{trapomega0chain}
is simply
\begin{eqnarray}
\Omega^{[0]}_T  \opsimeq_{T \to +\infty}e^{\displaystyle  T  S^{[0]}  } = e^{\displaystyle T \ln 2} = 2^T
 \label{trapomega0direct}
\end{eqnarray}
as it should.



\end{document}